\documentclass[a4paper,12pt]{article}
\usepackage[utf8]{inputenc}
\usepackage{cancel}
\usepackage{ulem}
\usepackage{amsfonts}
\usepackage{amssymb}
\usepackage{graphicx}
\usepackage{amsmath}
\usepackage{enumerate}
\usepackage{mathtools}
\usepackage{subfig}
\usepackage{color}
\usepackage{tikz}\usetikzlibrary{calc}
\usepackage{float}
\usepackage{here}
\usepackage{cite}
\usepackage{mathrsfs}
\usepackage{float,epsfig}
\usepackage{dcolumn}
\usepackage{tabularx}
\usepackage{multirow}
\usepackage{graphicx}
\usepackage{bm}
\usepackage{amsmath,amssymb,amsthm}
\usepackage[colorlinks=true,linkcolor=blue,citecolor=red]{hyperref}
\newcommand{\be}{\begin{equation}}
\newcommand{\ee}{\end{equation}}
\newcommand{\bea}{\setlength\arraycolsep{2pt} \begin{eqnarray}}
\newcommand{\eea}{\end{eqnarray}}

\def\0{{\sst{(0)}}}
\def\1{{\sst{(1)}}}
\def\2{{\sst{(2)}}}
\def\3{{\sst{(3)}}}
\def\4{{\sst{(4)}}}
\def\5{{\sst{(5)}}}
\def\6{{\sst{(6)}}}
\def\7{{\sst{(7)}}}
\def\8{{\sst{(8)}}}
\def\sst#1{{\scriptscriptstyle #1}}

\makeatletter \@addtoreset{equation}{section}

\usepackage{multirow}
\usepackage{tikz}

\begin{document}
%

\title{\normalsize
{\bf \Large	  Black Holes and Black Strings  in  M-theory on Calabi-Yau threefolds with four    K\"{a}hler parameters }}
\author{ \small A. Belhaj$^1$, H. Belmahi$^1$\thanks{Corresponding authors: hajar-belmahi@um5.ac.ma, abderrahim.bouhouch@um5r.ac.ma }, A. Bouhouch$^{1}$,  S. E. Ennadifi$^2$, M. B. Sedra$^3$ 
\footnote{Authors in alphabetical order.}
	\hspace*{-8pt} \\
	{\small  $^1$ Faculty of Science, Mohammed V University in Rabat,  4 Avenue Ibn Battouta,   Rabat, Morocco }\\
	{\small  $^2$  LHEP-MS,   Mohammed V University in Rabat, Rabat, Morocco}\\
	{\small  $^3$ LPMS, University of Ibn Tofail, Kénitra, Morocco}
 }

 \maketitle

	\begin{abstract}
{\noindent}   
Combining  toric geometry techniques and $\mathcal{N}=2$ supergravity formalisms, we study 5D black branes in the 
M-theory compactification on a four parameter Calabi-Yau threefold.  First, we
investigate  5D BPS and non-BPS black holes  that are   derived  by 
wrapping M2-branes on non-holomorphic 2-cycles in such a  toric  Calabi-Yau manifold.   Concretely, we  provide the allowed electric charge regions of BPS and non-BPS black hole states that are obtained by surrounding M2-branes over  appropriate 2-cycles. Then,  we   approach the black hole thermodynamic  behavior by computing the entropy and the temperature.  By evaluating the recombination factor, we examine the  stability of such  non-BPS black holes.  Precisely, we find stable and unstable solutions depending on  the allowed electric charge regions. After that, we   study  5D black strings   by 
wrapping  M5-branes on non-holomorphic dual  4-cycles in the proposed  toric   Calabi-Yau  manifold  by focusing on  the stability behaviors.  In the allowed regions of   the moduli space of  the non-BPS stringy  solutions, we find  stable and unstable states  depending on   the   magnetic charge values.

\end{abstract}
 \newpage
\tableofcontents

%

\newpage

\newpage

\section{Introduction}

Recently, a special interest has been developed to  the study of  charged black holes and black strings in  5 dimensions (5D).  These black objects have been  built  from 
supersymmetric M-theory on  Calabi-Yau   (CY)  threefolds  using the compactification mechanism \cite{1,2,3,4,5,6}. In  this approach, the BPS and the non-BPS black
   states have been derived by  help of  the 5D $\mathcal{N}=2$
supergravity formalism elaborated in the investigation  of lower dimensional supersymmetric  models \cite{7,70,71,72}.   A close examination shows that   such  charged   black holes and black strings  arise by  wrapping    M2  and M5 
branes  on  non-holomorphic 2-cycles and   dual 4-cycles   of the CY threefolds via intersecting  number  computations, respectively.   These cycle behaviors   are controlled by a
real number $h^{1,1}$ being  the K\"{a}hler moduli
space dimension.  The latter is  needed not  only to  specify the corresponding 5D spectrum fields  that are
derived  from  the M-theory compactification but also to determine the effective potential of   the BPS and the non-BPS  solutions in 5D.   Two and three  dimensional  K\"{a}hler moduli spaces  have been  investigated\cite{1,2,3,4,5,6}.   For two dimensions, for  instance, concrete models  have been studied by proposing   two   parameter complete intersection Calabi-Yau   (CICY)  models  in  the ordinary projective spaces   where  the stability behaviors  of such black brane
objects have been discussed. These studies have been  conducted  by calculating a  scalar quantity
called the recombination factor $R$.   Stable and unstable black  object 
solutions have been obtained   corresponding  to   $R<1$ and $R>1$, respectively \cite {1}. Later,  CICY models  in the  weighted  projective spaces have been  proposed by  insisting on the  weight value effect on the stability and the  thermal behaviors of  certain 5D black  M-branes   including black holes and   black strings\cite{4,5}. 

More recently,  these works  have been extended to three  parameter   CY threefolds in M-theory scenarios.  Two different CY descriptions  have been elaborated.   Concretly,  a toric description   of M-theory scenarios   has  been proposed   by elaborating a generic discussion. Precise computations   for   CY threefolds regarded    as  hypersurfaces in toric varieties (THCY)     have been  provided  for  $h^{1,1}=3$. In this way,   5D
BPS and   non-BPS black brane  configurations involving stable and unstable
behaviors have been obtained   using numerical computations \cite{3}. Alternatively,   a
three parameter model of   CICY     dealing with  such  black 
brane stability  behaviors based on  the $\mathcal{N}=2$ 5D 
formalism has been  investigated \cite{6}. Precisely,  a  M-theory CY threefold  in  the 
 $\mathbb{P}^{1}\times \mathbb{P}^{1}\times \mathbb{P}^{2}$ projective space product  has been studied by  calculating  the corresponding  effective potential.  In this regard,    5D BPS and non-BPS black  brane  solutions have been derived and examined.   
 Stable and unstable black branes  depending   on  the charge
regions of the   K\"{a}hler moduli
space have been determined using analytical and numerical computations.

 In this work, we study 5D black branes in the 
M-theory compactification  on a four parameter THCY by combining  toric geometry techniques and $\mathcal{N}=2$ supergravity formalisms.    First, we
investigate   5D BPS and non-BPS black holes  that are obtained by 
wrapping  M2-branes on non-holomorphic 2-cycles in such a toric  CY manifold.   Concretely, we determine  the  allowed  electric charge regions of  the  black hole  moduli space and  approach certain thermodynamic behaviors by  computing the corresponding quantities including the temperature and the entropy.
  Then,  we inspect the   stability behaviors of such  5D  black holes via the calculation of the  recombination factor $R$. 
After that, we approach  5D black strings being  derived  by 
wrapping  M5-branes on non-holomorphic 4-cycles in  the proposed  CY toric geometry.    Calculating  the recombination factor $R$ for  5D black strings, we find
various stable and unstable solutions depending on the  magnetic charge
regions of the involved moduli space.

The organization of this paper is as follows. In section 2, we provide 
a  concise discussion on 5D M-theory  black  branes  using $\mathcal{N}=2$ supergravity formalisms.  In
section 3, we investigate  5D BPS and non-BPS black holes
by  wrapping M2-branes on  2-cycles  inside a    CY  threefold  with four    K\"{a}hler parameters using toric geometry techniques. In section 4,
we move to study    the 5D BPS and the  non-BPS black strings  by  wrapping M5-branes on  4-cycles  in such a  toric CY  threefold. The last section is devoted to certain  final remarks and  open issues.

\section{M-theory black branes on  CY geometries}
To start, we would like to provide a  concise discussion   on  5D black  branes from   the M-theory compactification  using  $\mathcal{N}=2$  supergravity techniques  elaborated in the study of stringy  spectrums in lower dimensional   space-times.  Precisely, these solutions 
can be derived from the compactifaction scenario on  CY threefolds needed  to engineer models   involving  minimal supercharges in 5D \cite{8,9,10,11,12,13,14,15,16}. Alternatively, these non-trivial geometries  have been  also exploited 
 in the compactification of  type II superstrings producing  4D  $\mathcal{N}=2$   
supersymmetric classes based on mirror symmetry tools \cite{17,18,19}.    It is  worth recalling that,  at low energies, M-theory represents an  eleven-dimensional supergravity model   \cite{91}. 
This  can generate  certain  non-perturbative limits of superstring theories  via the compactification mechanism  on specific
specific geometries  with non-trivial holonomy groups.
In the  M-theory context,  the black brane objects can be built by considering   M2 and
M5 solitonic solutions living in 11D.   With  help of    the  compactification mechanism, these black  objects can be approached  using the  5D $\mathcal{N}=2$ supergravity formalisms. The associated physics depends   on the  CY   moduli   spaces  coordinated by  the  K\"{a}hler and the complex  deformation parameters  associated with  the Hodge numbers  $h^{1,1}$ and    $h^{2,1}$, respectively \cite{160,161,162,163,164,165,166,167,168}.  Forgetting about hypermultiplet spectrums linked  to  the Hodge number $h^{2,1}$,    the  black  object physics  in such  M-theory compactifications   can  be  studied  using  the following 5D   Maxwell-Einstein action 
\begin{equation}
S=\frac{1}{2\kappa _{5}^{2}}\int d^{5}x\left( R\star \mathbb{I}%
-G_{IJ}dt^{I}\wedge \star dt^{J}-G_{IJ}F^{I}\wedge \star F^{J}-\frac{1}{6}%
C_{IJK}F^{I}\wedge F^{J}\wedge A^{K}\right) 
\end{equation}%
describing   the dynamics of $h^{1,1}$ vector multiples indexed by $I$ involving  the  scalar   K\"{a}hler  moduli   $t_{I}$  and gauge fields $F^{I}=dA^{I}$.    The symmetric tensor $G_{IJ}$ indicates 
the moduli space metric being a relevant  piece  to compute   the
effective scalar potential needed to examine the associated stability  behaviors.  The  tensor elements   $C_{IJK}$  provide the intersecting numbers which  could
be fixed once the CY  manifolds  are  built. 
 5D black holes, for instances, carry  $q^{I}$
electric charges under  the  $U(1)^{\otimes h^{1,1}}$   abelian gauge  symmetries. These  electric  charges have been exploited   to calculate the corresponding  effective potential  via 
the moduli space metric  $G_{IJ}$. It has been suggested  that this effective  potential  can take the following form  
\begin{equation}
V_{eff}^{e}=G^{IJ}q_{I}q_{J},  \qquad I,J=1,\ldots, h^{1,1}.\label{Vbh}
\end{equation}
This scalar potential is invariant under a $Z_2$ symmetry acting as follows
\begin{equation}
Z_2:  q_{I} \to -q_{I}
\end{equation}
which could be exploited in certain discussions corresponding to the  allowed electric charge regions.  For dual black strings with 
magnetic charges $p^{I}$,  however, the effective scalar potential   can be expressed as 
\begin{equation}
V_{eff}^{m}=4G_{IJ}p^{I}p^{J}.  \label{vm}
\end{equation}%
In this  context,  $G_{IJ}$ can be written in
terms of the volume $ \mathcal{V}$  of  the involved  CY threefolds  via the relation
\begin{equation}
G_{IJ}=-\frac{1}{2}\partial _{I}\partial _{J}\log (\mathcal{V}),
\end{equation}%
where  one has used 
\begin{equation}
 \label{Vol}
\mathcal{V}=\frac{1}{3!}C_{IJK}t^{I}t^{J}t^{K}.
\end{equation}
In  M-theory scenarios, certain models of two  and three  parameter CY threefolds have been
studied using at least two  classes of CY spaces relaying on  THCY  and  CICY   descriptions  \cite{1,2,3,4,5,6}.   Using  analytical and numerical computations for    the recombination factor $R$,   stable and non-stable solutions 
for certain  5D  black holes and  black strings have   been examined. They are associated  with   $R<1$ and $R>1$ conditions, respectively\cite{3}. 
 
 In what  follows, we would like to extend  such  works to   four   parameter CY threefolds using a toric description  in  the M-theory compactification mechanism by help  of  M2 and M5-branes.

\section{ M-theory black holes  from  a  THCY with   $ h^{1,1}=4$ }
In this section, we would like  to  deal with   5D black hole  behaviors  from a  four parameter CY threefold in  the M-theory compactification scenarios.   This manifold  can be constructed using   toric   geometry techniques \cite{20,21,22,23,24,25,26} . These  mathematical tools have  been  largely  used  in  the geometric engineering method  being   exploited  to provide lower dimensional gauge models from different roads  including   type II superstrings,  M-theory on G2 manifolds,  and  F-theory with elliptic compactifications \cite{28,280,29}. Roughly speaking, a toric manifold is associated with  a polytope  encoding the geometric  data  of the associated  CY manifolds defined as hypersurfaces.  In this way,  this  manifold  can be   described  in terms of   quotient spaces generating the ordinary and  the weighted projective  spaces.    To construct such quotient  spaces,  one needs, usually,  a $n+r$ dimensional complex space  ${\mathbb{C}^{n+r}}$  coordinated  by $z_1,\ldots,   z_{n+r}$ with  $r$ ${\mathbb{C}^*}$  scale symmetries    acting as follows 
  \begin {equation}
\label{toric}
{\mathbb{C}^*}^a: \;  z_i   \to \lambda^{q_i^a} z_i,\;\; i=1,2,\ldots, n+r,  \quad a=1,
  2,\ldots,r,
  \end{equation}  
  where   $\lambda$ is a  non-zero complex number and  $q_{i}^{a}$ are integers where each $a$  generates 
the so-called  Mori vector in toric  geometry language\cite{30,31}.  In fact, they    generalize  the  weight vector
of the  complex  $n$-dimensional  weighted projective space
 ${\bf W\mathbb{C}\mathbb{P}}^n{(w_1, \ldots, w_{n+1})}$  associated with one Mori  vector $w_i=q_i$.  Roughly speaking, 
 a general toric variety  ${{\cal \bf V}^n}$ can be defined  by the following  symplectic
quotient space 
  \begin {equation}
  {{\cal \bf V}^n} = {\mathbb{C}^{n+r}\setminus U\over {\mathbb{C}^*}^r},
  \end{equation}
     where  $U$ is considered as  a subset  of
  $\mathbb{C}^{k}$ required by triangulation  configurations.  A close examination shows that one  can regard 
   ${{\cal \bf V}^n}$  as a    $ T^n$  fibration. This can be understood 
    by dividing   $ T^{n+r}$  by the  $U(1)^r$  gauge symmetry
   \begin {equation}
    z_i\to e^{iq^a_i \theta^a} z_i,\quad a=1,\ldots,r,
      \end{equation}
   where $\theta^a$ are the generators of the  $ U(1)$ phase symmetries. In this way,       ${{\cal \bf V}^n}$    can 
 be  represented by a  toric
     graph  $ \Delta({{\cal \bf V}^n})$ spanned by  $ k=n+r$
     vertices $ v_i$ in the   $\bf Z^n$ lattice satisfying the following toric relations
      \begin {equation}
    \sum \limits _{i=1}^{n+r} q_i^av_ i=0,\quad  a=1,\ldots,r.
  \end{equation}
  It turns out that 
such a toric  manifold   exhibits  a beautiful  physical realization  via  ${\cal N}=2$ linear sigma models  in two dimensions involving  the 
$U(1)^r$ gauge  fields coupled to   $n+r$ chiral fields $\phi_i$   with a    
matrix gauge charge  $q_i^a$  ~\cite{32}. Up to  the $U(1)^r$ gauge
transformations,   ${{\cal \bf V}^n}$  can be considered as  a solution  of 
  the D-term  flatness condition 
\begin {equation}
  \label{MinimumDTerm}
  \sum \limits _{i=1}^{n+r} q_i^a|\phi_i|^2=\rho^a,
\end{equation}
where  the  quantities  $\rho^a$  represent  the  Fayet-Iliopoulos (FI) coupling parameters~\cite{32,33}.    In  toric  variety  ${{\cal \bf V}^n}$,   
a hypersurface  with a  vanishing first Chern class  provides  a  $(n-1)$  dimensional CY  manifold.  Using   toric  techniques, CY threefolds have been classified by Kreuzer and Sakrke\cite{21}. The present investigation concerns   a CY threefold  encoded in Krenzer-Skarke database  involving  all geometrical  data of all different CY  manifolds in terms of the ambient  toric polytope information \cite{21,22,23,24,25,26}.   Supported  by such activities,  we consider   a four parameter  CY threefold with the following toric data 
\begin{center}
\begin{tabular}{|p{1.0cm}|p{2.5cm}|p{3.4cm}|p{6.3cm}|p{2.4cm}|} 
\hline
  
Poly tope & Weight matrix &  K\"{a}hler  cone ma trix & Mori cone matrix &($ h^{1,1} $, $ h^{1,2} $, $ \chi$  )\\  
\hline
1310&$\begin{array}{rl}
\left(\begin{array}{cccc}
  0 & 0 &0 &1  \\
 0 &0 & 1 &0  \\ 
 0 &1 & 1& 0 \\
 1 &1 & 2& 0 \\
 0 &1 & 0& 0 \\
 1 &1 & 2& 1 \\
 3 &4 & 6& 2 \\
 1 &0 & 0& 0
\end{array}\right)\end{array}$
 & $\begin{array}{rl}\left(\begin{array}{cccc}
 0 & -1 & 0  &0  \\ 
 0 & 1  &  0 &1  \\ 
  0&  0 &  1 & 0 \\
  1&  0 & 0  & -2\\
\end{array}\right)\end{array} $& $\left(\begin{array}{cccccccc}
 -1 & 1  & 0  &1 &-1  & 0  &0 &0  \\ 
  0 &- 1  & 0  &0 &1  & 0  &1 &1  \\ 
 1 & 0 & 0  &0 &0  & 1 &2 &0  \\
 0 & 1  & 1  &0 &0  & 0  &0 &-2  \\
\end{array}\right)
$&  (4,106,-204)\\
\hline
\end{tabular}
\end{center}
 A close inspection,   in the study of  the black holes and the  black strings  in  the M-theory compactification  on CY threefolds,  reveals that the relevant  data   are the intersection numbers. These  geometric  quantities are needed to calculate   the effective potential of   the black  brane objects including the black holes and the black strings\cite{1}. For  the present  toric CY  threefold,    we should  compute the tensor elements   
$ C_{IJK} $ with $ I,J,K=1, \ldots,4 $ taking the  following matrix form 
\begin{eqnarray}
C_{1JK}=\left(\begin{array}{cccc}
C_{111}& C_{112} & C_{113} & C_{114} \\ 
C_{121} & C_{122} & C_{23}& C_{124} \\ 
C_{131} & C_{132}  & C_{133} & C_{134}  \\ 
C_{141}  & C_{142}  & C_{143}  & C_{144} 
\end{array}\right), \,\,\,\,
C_{2JK}=\left(\begin{array}{cccc}
C_{211}& C_{212} & C_{213} & C_{214} \\ 
C_{221} & C_{222} & C_{223}& C_{224} \\ 
C_{231} & C_{232}  & C_{233} & C_{234}  \\ 
C_{241}  & C_{242}  & C_{243}  & C_{244} 
\end{array}\right) 
\\
C_{3JK}=\left(\begin{array}{cccc}
C_{311}& C_{312} & C_{313} & C_{314} \\ 
C_{321}& C_{322}& C_{323}& C_{324} \\ 
C_{331}& C_{332}& C_{333}& C_{334}  \\ 
C_{341}& C_{342}& C_{343}& C_{344} 
\end{array}\right), \,\,\,\,
C_{4JK}=\left(\begin{array}{cccc}
C_{411}& C_{412} & C_{413}& C_{414} \\ 
C_{421}& C_{422}& C_{423}& C_{424} \\ 
C_{431}& C_{432}& C_{433}& C_{434}  \\ 
C_{441}& C_{442}& C_{443}& C_{444}
\end{array}\right).
\end{eqnarray}
Using  the equation given in  (\ref{Vol}), appropriate calculations provide the following volume expression of the proposed  toric  CY threefold
\begin{equation}
\mathcal{V}=\frac{1}{3} \left(3 t_1 t_{3}(t_{3}+2 t_4)+t_{2}^3-3 t_{2}^{2} t_3+3t_{2} t_{3}^2+t_{3}^3-6 t_{3} t_{4}^2\right).
\end{equation}
Exploiting now the relation (\ref{Vbh}), we can   establish    the effective potential  of 5D black holes in such M-theory scenarios. Precisely, it takes    the form 
\begin{equation}
V_{eff}^{BH}(q_I,t_I)=\frac{G(q_I,t_I)}{T(t_I)}
\end{equation}
where  $T$  is  a  geometric function  which reads as  
\begin{equation}
T(t_I)=  3 \left(t_1( t_3+2 t_4)-t_{2}^2+t_2 t_3-2 t_4^2\right).
\end{equation}
The scalar  quantity  $G(q_I,t_I)$  is found to  be 
\begin{equation}
G(q_I,t_I)=
g^{IJ}(t_I)q_{I}q_{J}
\end{equation}
where one has used  the following matrix elements 
\begin{align*}
g^{11}&=6(t_3+2 t_4) t_1^3+(-6 t_2^2+6 t_3 t_2+7 t_3^2-36 t_4^2) t_1^2+2(7 t_3^3+10 t_4 t_3^2-12 t_2 t_4 t_3+12 t_4 (t_2^2+2 t_4^2)) t_1\\
&-2(3 t_2^4-6 t_3 t_2^3+8 t_3^2 t_2^2+12 t_4^2 t_2^2-5 t_3^3 t_2-12 t_3 t_4^2 t_2-2 t_3^4+12 t_4^4+10 t_3^2 t_4^2)
\,\\
g^{12}&=-3(2 t_2-t_3)(t_3+2 t_4) t_1^2+(6 t_2^3-9 t_3 t_2^2-4 (t_3^2+3 t_4 t_3-3 t_4^2) t_2+t_3(7 t_3^2+12 t_4 t_3-6 t_4^2)) t_1\\
&+2 (t_2-t_3) t_3(3 t_2^2-3 t_3 t_2-t_3^2+6 t_4^2)
\,\\
g^{13}&=t_3^2 \left(-6 t_2^2+6 t_3 t_2+2 t_3^2-12 t_4^2+7 t_1 t_3+12 t_1 t_4\right)
\,\\
g^{14}&=-3 t_2^4+6 t_3 t_2^3+(3 t_1(t_3+2 t_4)-2 (4 t_3^2+3 t_4 t_3+6 t_4^2)) t_2^2+t_3(5 t_3^2+6 t_4 t_3+12 t_4^2\\
&-3 t_1 (t_3+2 t_4)) t_2+t_1(6 t_3^3+17 t_4 t_3^2+18 t_4^2 t_3+12 t_4^3)+2(t_3^4+t_4 t_3^3-5 t_4^2 t_3^2-6 t_4^3 t_3-6 t_4^4)
\,\\
g^{22}&=-3 t_2^4+6 t_3 t_2^3-6 t_3^2 t_2^2+2 t_3^3 t_2+t_3^4+12 t_4^4-8 t_3^2 t_4^2+3 t_1^2(t_3+2 t_4)^2+4 t_1 (t_3^3+2 t_4 t_3^2-3 t_4^2 t_3-6 t_4^3)
\,\\
g^{23}&=t_3^2 \left(-3 t_2^2+2 t_3 t_2+t_3^2-6 t_4^2+3 t_1 \left(t_3+2 t_4\right)\right)
\,\\
g^{24}&=3(t_3+2 t_4) t_2^3-3 t_3 (2 t_3+3 t_4) t_2^2+(2(t_3^3+t_4 t_3^2+3 t_4^2 t_3+6 t_4^3)-3 t_1 (t_3+2 t_4)^2) t_2\\
&+t_3 (t_3+t_4)(t_3^2-6 t_4^2+3 t_1(t_3+2 t_4))
\,\\
g^{33}&=t_3^2 \left(-6 t_2^2+6 t_3 t_2+t_3^2-12 t_4^2+6 t_1 \left(t_3+2 t_4\right)\right)
\,\\
g^{34}&=t_3^2 (-3 t_2^2+3 t_3 t_2+t_3^2-6 t_4^2+t_3 t_4+3 t_1 (t_3+2 t_4))
\,\\
g^{44}&=t_3^4+2 t_4 t_3^3-5 t_4^2 t_3^2-12 t_4^3 t_3+3 t_2 (t_3^2+2 t_4 t_3+2 t_4^2) t_3-12 t_4^4-3 t_2^2 (t_3^2+2 t_4 t_3+2 t_4^2)\\
&+3 t_1 (t_3^3+4 t_4 t_3^2+6 t_4^2 t_3+4 t_4^3).
\end{align*}

Having computed the black hole effective  scalar potential, we move now   to examine the  BPS and non-BPS  black hole   solutions in the proposed M-theory compactification followed by a stability analysis.

\subsection{BPS and non-BPS  black hole   solutions}
Here, we investigate  the  BPS and  the non-BPS  black hole behaviors with four charges  $q_I$   associated with $U(1) \times  U(1) \times  U(1) \times  U(1)$  gauge
fields  derived by   wrapping   M2-branes on 2-cycles in the   above four  parameter CY  toric geometry.  To do so, we  first elaborate   the BPS black hole solutions. To get such solutions, we  need  to  solve  the constraints 
\begin{equation}
q_I-2 \tau_I Z_e=0,\,\,\,I=1,2,3,4
\end{equation}
where  one  has used 
\begin{equation}
\tau_I= \frac{1}{2}  C_{IJK}    t^J t^K , \qquad  Z_{e}=q_1 t_1+ q_2 t_2+ q_3 t_3+q_4 t_4.
\end{equation}
To handle such equations,   the  local coordinates of the black hole  moduli spaces should be used. More precisely, we consider  the  following inhomogeneous  variables 
\begin{equation}
 x=\frac{t_{1}}{t_{4}}, \qquad y=\frac{t_{2}}{t_{4}}, \qquad z=\frac{t_{3}}{t_{4}},  \qquad  \alpha=\frac{q_{1}}{q_{4}}, \qquad \beta=\frac{q_{2}}{q_{4}}, \qquad \gamma=\frac{q_{3}}{q_{4}}
\end{equation}
where $ q_I$ and $t_I$  represent  the homogeneous  charges and  the geometric   variables  of  the  5D black hole  moduli space, respectively.  In terms of the local coordinates, the above constraints provide the following three equations 
\begin{equation}
\begin{array}{rl}
-\frac{q_4(-2 \alpha  (x-2)+z+2)}{2 (x-2)} &= 0 \\
\frac{q_4(2 \beta  (x-2)+2 y-z)}{2 (x-2)}&=0\\
  \frac{q_4 \left(2 x ((\gamma -1) z-1)+y^2-2 y z-z^2-4 \gamma  z+2\right)}{2 (x-2) z}&=0.
\end{array}
\end{equation}
Solving this system of equations,  we  can get  the local charge variables in terms of the local coordinates of the  K\"{a}hler  moduli in the following form
\begin{equation}
\alpha = \frac{z+2}{2 (x-2)},\,\,\beta = \frac{z-2 y}{2 (x-2)},\,\,\gamma= \frac{2 x (z+1)-y^2+2 y z+z^2-2}{2 (x-2) z}.
\label{bbs}
\end{equation}
Indeed, it is possible to express to the  local geometric coordinates   as functions  of the local charge variables. Two triplet solutions are obtained $(x_+, y_+,  z_+),  $  and  $(x_-, y_-, z_- )$  which are given by 
\begin{eqnarray}
x_{\mp}&=&\frac{14 \alpha ^3\mp\xi( \alpha,\beta,\gamma)+\alpha ^2 (-4 \beta -8 \gamma +11)-\alpha  \left(2 \beta ^2+\beta +2 \gamma -1\right)}{\alpha  \left(7 \alpha ^2-2 \alpha  (\beta +2 \gamma -2)-\beta ^2\right)} \nonumber\\
y_{\mp}&=& \frac{-4 \alpha ^3\mp\alpha  \left(\xi( \alpha,\beta,\gamma)-2 \beta ^2-2 \beta  \gamma +\beta \right)\pm\beta  \xi( \alpha,\beta,\gamma)+\alpha ^2 (-2 \beta +2 \gamma -3)}{\alpha  \left(7 \alpha ^2-2 \alpha  (\beta +2 \gamma -2)-\beta ^2\right)} \nonumber\\
z_{\mp}&=&\frac{\mp2\xi( \alpha,\beta,\gamma)-8 \alpha ^2+2 \alpha  (\beta +2 \gamma -3)+2 \beta ^2}{7 \alpha ^2-2 \alpha  (\beta +2 \gamma -2)-\beta ^2}
\end{eqnarray}
where one has used  \begin{eqnarray}
\xi( \alpha,\beta,\gamma)=\sqrt{\alpha ^2 \left(2 \alpha ^2+\alpha  (-4 \beta -8 \gamma +2)+2 \beta ^2+\beta  (4 \gamma -2)+(1-2 \gamma )^2\right)}. \end{eqnarray}
To determine the possible regions  of  the allowed  electric charges,  strong constraints should  be imposed on the geometric local variables of the black hole moduli space to ensure that  they lie within the K\"{a}hler cone.  A close examination shows that each triplet  solution  should be analyzed separately. Notably, the  first  triplet $(x_-, y_-, z_- )$ reveals larger regions of the  allowed  electric charges. Similar behaviors can be elaborated for the second triplet $(x_+, y_+, z_+)$.  As illustrated in Fig.(\ref{F1}), the allowed  electric charge regions for BPS black holes in the  $(\alpha,\beta)$ plane  are shown for the $(x_-, y_-, z_- )$  triplet by taking certain  values of the local  electric charge $\gamma$.

\begin{figure}[h!]
\label{F1}
\begin{center}
	\includegraphics[scale=1]{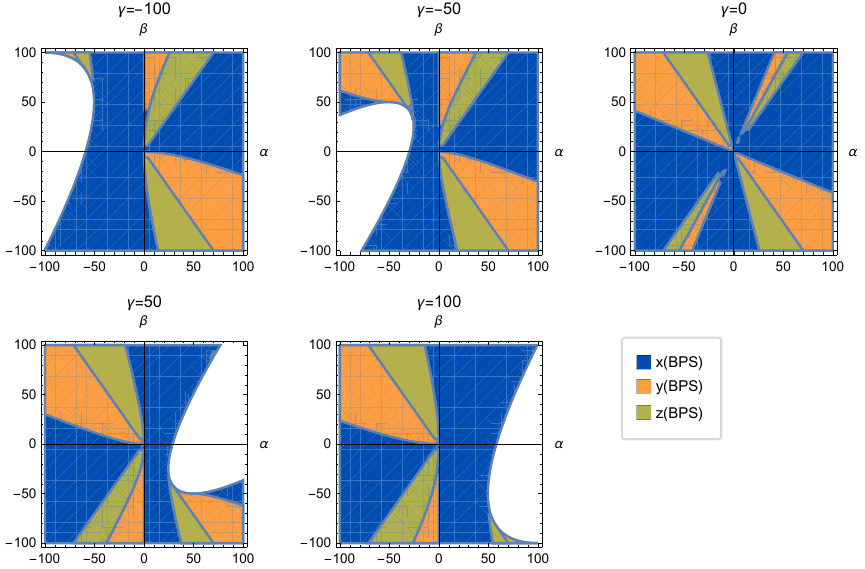}
	\caption{{\it \footnotesize  Electric charge regions for   BPS black hole sates associated with  the second triplet $(x_-, y_-, z_- )$.  }}
\label{F1}
\end{center}
\end{figure}
In this figure, the white regions indicate the absence of large black hole solutions. It has been observed that for each fixed value of $\gamma$, the configurations provide only half-cone black hole solutions. Interestingly, the negative and  the positive values of 
$\gamma$ can be combined to recover the symmetric cone of the BPS black holes appearing in two parameter CY models\cite{1,4,5}. Moreover, in the special case of 
$\gamma=0$, the cone symmetry is fully restored.

According to  \cite{4,5,6}, we approach certain thermodynamic behaviors of such  black hole solutions. In particular,    we  can calculate the  entropy of the BPS black holes   via the relation 
\begin{equation}
S_{BPS}=\frac{ 2 \pi}{3\sqrt{3}} |Z_e|^{3/2}.
\end{equation}
Indeed, this entropy   is found to be 
\begin{equation}
S_{BPS}=\frac{ 2 \pi}{3\sqrt{3}}| t_4 q_4|^{3/2} |x\alpha+y\beta+z\gamma+1|^{3/2}.
\end{equation}	
Using  the CY threefold volume,  $t_4$   can be  expressed  in terms of  the local variables of the  black  hole moduli space  as follows
\begin{equation}
t_4=\frac{1}{2^{2/3} \sqrt[3]{(x-2) z (\alpha  x+\beta  y+\gamma  z+1)}}.
\end{equation} 
Computations   show that the  BPS black hole  entropy  can take the  form 
\begin{equation}
S_{BPS}=\frac{1}{9} \sqrt{2} \pi  \left|\frac{{q_4} (\alpha  x+\beta  y+\gamma  z+1)}{\sqrt[3]{(x-2) z (\alpha  x+\beta  y+\gamma  z+1)}}\right|^{3/2}.
\end{equation}
Examining the cubic root charge behaviors of this expression provides a pathway to explore other thermodynamic quantities. Notably, numerous extended forms of  the entropy have been introduced in various  black hole studies \cite{OS1,OS2}. Inspired by  such generalized entropies, we can investigate the thermodynamic behaviors of these 5D black hole solutions, with particular attention to  the temperature as a key parameter. Specifically, we could identify the obtained entropy $S_{\text{BPS}}$ with the Brown one, denoted as $S_{B}$
\begin{equation}
S_{B}=\left(\frac{A}{A_{p_{1}}}\right)^{3/2},
\end{equation}
where    one has used  $ A_{p_{1}}=4G $ being  the Planck area and $ A=4\pi r_{h}^{2} $.    In this way,  $\Delta$    indicates a    deformed   quantum gravity     dimensionless  parameter.   Considering  a   maximal quantum deformation  required   by   $\Delta=1$,  we find     the event horizon radius  
\begin{equation}
r_{h}=\frac{1}{2}\left(\frac{1}{9} \sqrt{2} \pi\right)^{1/3} \left|\frac{{q_4} (\alpha  x+\beta  y+\gamma  z+1)}{\sqrt[3]{(x-2) z (\alpha  x+\beta  y+\gamma  z+1)}}\right|^{1/2}.
\end{equation}
Exploiting the thermodynamic  techniques,  we   obtain   the   Hawking  temperature
\begin{equation}
T_{H}=\left(\frac{9}{\sqrt{2} \pi} \right)^{1/3}\left|\frac{\sqrt[3]{(x-2) z (\alpha  x+\beta  y+\gamma  z+1)}}{{q_4} (\alpha  x+\beta  y+\gamma  z+1)}\right|^{1/2}.
\end{equation}
For generic regions of the moduli space,   such a  temperature exhibits inverse  square root  charge behaviors  matching with the results obtained   in the literature. 
 
	 

To obtain  5D    non-BPS  black hole solutions, we can exploit  the Lagrange multipliers  used  in  the elaboration of  the equations of motion in dynamical systems. Concretely,  this could be approached  
via the relation
\begin{equation}g_{\cal V}=3 t_1 t_{3}(t_{3}+2 t_4)+t_{2}^3-3 t_{2}^{2} t_3+3t_{2} t_{3}^2+t_{3}^3-6 t_{3} t_{4}^2. \end{equation}
Solving  the  equations
\begin{equation}
\label{defVeff}
\dfrac{D_{I}V^e_{eff}}{D_{J}V^e_{eff}}=\dfrac{D_{I}g_{\cal V}}{D_{J}g_{\cal V}},\,\,\,\,\,I,J=1,2,3,4
\end{equation}
one can find the possible solutions, where  one has used $ D_{I}= \partial_{I}-\frac{2}{3\nu}\tau_I$. Instead of giving large equations, we give, however,  only  the possible accessible  solutions.   Indeed, computations for building such  models reveal  that   we have   five solutions which are listed  as follows:\\
Solution 1
\begin{eqnarray}
\label{1}
\alpha &=&\frac{z+2}{2 (x-2)} \nonumber\\
\beta &= &\frac{z-2 y}{2 (x-2)} \\
\gamma &=& \frac{2 x (z+1)-y^2+2 y z+z^2-2}{2 (x-2) z}. \nonumber
\end{eqnarray}
Solution 2
\begin{eqnarray}
\label{2}
  \alpha &= &-\frac{3 (z+2)^2}{2 \left(3 x (z+2)-6 y^2+6 y z+2 z (z+3)\right)} \nonumber\\
  \beta& = &\frac{3 (z+2) (2 y-z)}{2 \left(3 x (z+2)-6 y^2+6 y z+2 z (z+3)\right)}\\
  \gamma&=&\frac{6 x \left(z^2+3 z+2\right)-3 y^2 (3 z+2)+6 y z^2+z^3-2 z^2-18 z-12}{2 z \left(3 x (z+2)-6 y^2+6 y z+2 z (z+3)\right)}.\nonumber
\end{eqnarray}
Solution 3
\begin{eqnarray}
\label{3}
 \alpha &=& -\frac{(z+2)^2 \left(12 x (z+2)-12 y^2+12 y z+5 z^2-24\right)}{\Lambda(x,y,z)}  \nonumber\\
  \beta&=& -\frac{(z+2) (z-2 y) \left(12 x (z+2)-12 y^2+12 y z+5 z^2-24\right)}{\Lambda(x,y,z)}\\
 \gamma &=& -\frac{\varsigma(x,y,z)}{\Lambda(x,y,z)} \nonumber
\end{eqnarray}
where one has used 
{\small 
\begin{align*}
\varsigma(x,y,z)&=24 x^2 (z+2)^2+2 x (z+2) \left(6 y^2 (z-4)+36 y z+z^3+11 z^2+12 z-48\right)-12 y^4 (z-2)\\
&+12 y^3 (z-6) z+y^2 \left(3 z^3+26 z^2-48 z+96\right)+2 y \left(z^3+16 z^2+12 z-72\right) z+z^5+6 z^4\\
&+6 z^3-44 z^2-48 z+96\\ \,\\
\Lambda(x,y,z)&=24 x^2 (z+2)^2+6 x (z+2) \left(-12 y^2+12 y z+3 z^2+8 z-8\right)+4 (12 y^4-24 y^3 z\\
&+y^2 (5 z^2-12 z+24)+y z(7 z^2+12 z-24)+z (z^3+5 z^2-4 z-24)).
\end{align*}}
Solution 4
\begin{eqnarray}
\label{4}
\alpha &=& \frac{z+2}{2 (x-2)} \nonumber \\
 \beta &= &\frac{z-2 y}{2 (x-2)} \\
 \gamma &=& \frac{\Gamma(x,y,z)}{2 (x-2)\Delta(x,y,z)} \nonumber
\end{eqnarray}
where the function $\Delta(x,y,z)$ is expressed as follows
{\small 
\begin{align*}
\Gamma(x,y,z)&=-24 x^3 (z+2)^2+2 x^2 (z+2) (6 y^2 (z+6)-24 y z+7 z^3+9 z^2+12 z+72)
\\&-x (24 y^4 (z+3)-24 y^3 (z+4) z+y^2 (23 z^3+60 z^2+96 z+288)-2 y (14 z^3+23 z^2\\
&+24 z+96) z-7 z^5-12 z^4+46 z^3+72 z^2+96 z+288)+12 y^6-24 y^5 z+3 y^4(7 z^2+24)\\
&-y^3 z(23 z^2+96)+4 y^2 (2 z^4+15 z^2+36)+y z(7 z^4-46 z^2-96)+z^6-12 z^4+36 z^2+96\\
\,\\
\Delta(x,y,z)&=z(12 x^2 (z+2)^2+x (z+2)(-24 y^2+24 y z+5 z^2-48)+12 y^4-24 y^3 z\\
&+y^2(7 z^2+48)+y z(5 z^2-48)+z^4-10 z^2+48).
\end{align*}}
Solution 5 
\begin{eqnarray}
\label{5}
\alpha &=&\frac{z+2}{2 (x-2)}  \nonumber\\
\beta&=&\frac{\phi_{\mp}(x,y,z)}{2 (x-2)^2\chi(x,y,z)}\\
\gamma&=&\frac{\pm\varphi(x,y,z)}{2 (x-2)^2\chi(x,y,z)} \nonumber
\end{eqnarray}
where one has used 
{\small
\begin{align*}
\phi_{\mp}(x,y,z)&=288 x(x^2-3 x+2) z^2+12(6 x^3-x^2-33 x+22) z^3+72 (x-2) y^3 (z+2)(2 x+z-2)\\
&-108(x-2) y^2 (z+2) z (2 x+z-2)+15 (x-2) z^5+66 (x-2) x z^4+288 (x-2) (x-1)^2 z\\
&\mp2 \sqrt{2}\delta(x,y,z)-6 (x-2) y (z+2)(24 x^2 (z+2)+2 x(5 z^2-48)-z^3+2 z^2-24 z+48)\\
\,\\
\chi(x,y,z)&=12 y^2(6 x (z+2)-5 z^2-12)-12 y z(6 x (z+2)+z^2-12)+z^2 (18 x (z+2)+7 z^2-36)\\&-72 y^4+144 y^3 z)\\
\,\\
\varphi(x,y,z)&=-72 (x-2) y^4 z^3 (x+z)+12 y^3(-12 (x^2-3 x+2) z^3\pm\sqrt{2}\delta(x,y,z)+9 (x-2) z^5+6 (x-2) x z^4)\\
&+2y(-288 z^3+720 x z^3-576 x^2 z^3+144 x^3 z^3+144 x z^4-216 x^2 z^4+72 x^3 z^4-96 z^5\\
&+144 x z^5-48 x^2 z^5+60 x z^6-30 x^2 z^6-9 x z^7+18 z^7\\ 
&\pm \sqrt{2}\delta(x,y,z)(12-12x-6xz+2z^{2})\\
&+z(336 z^3-840 x z^3+672 x^2 z^3-168 x^3 z^3-264 x z^4+396 x^2 z^4-132 x^3 z^4-76 z^5+114 x z^5\\
&+10 x^2 z^5-24 x^3 z^5+24 x z^6-12 x^2 z^6+2 z^7-x z^7\\
&\pm\sqrt{2}\delta(x,y,z)(-12+12x+6xz+2z^{2})\\
&-3 y^2 (-24 x^3 z^3 (z+2)-2 x^2 z^3(21 z^2+38 z-72)+x z^3 (z^3+84 z^2+372 z-96)+\\
&+2(-z^6-124 z^4+3 \sqrt{2} z\delta(x,y,z)))
\end{align*}}
with  $ \delta=\sqrt{(x-2)^2 z^4 (z+2) (2 x+z-2)(3 x (z+2)-3 y^2+3 y z+z^2-6)^2} $. It has been observed that the first solution  (\ref{1})  has been already  considered  in the elaboration  the BPS  charge solutions  given in  (\ref{bbs}).  However,  the  remaining ones   do not  verify  the  BPS  equations  giving the   non-BPS  solutions which will be dealt with in the forthcoming  discussions.  Considering  such non-BPS solutions,   we can express the local geometric variables  in terms of the local charge quantities.   For each solution, computations provide  two triplets.     Due to  higher orders of the local geometric coordinates  for  certain solutions,   we approach     the allowed charge regions using two different methods. First, we consider lower orders   appearing in the second and the third solutions.  Then, we reconsider the remaining  ones via   attractor  techniques.  Indeed, for  the second  non-BPS  black hole  solution (\ref{2}), we find
\begin{align}
\begin{cases}
x^{2}_{\mp}=\frac{14 \alpha ^4+\alpha ^3 (4 \beta +8 \gamma -3)+\alpha ^2 (2 \beta ^2+\beta  (9-8 \gamma )-8 \gamma ^2+18 \gamma -7)\mp 2 \beta ^2\upsilon(\alpha,\beta,\gamma)+\alpha(-4 \beta ^3+\beta ^2 (4-8 \gamma )+2 \beta\upsilon(\alpha,\beta,\gamma)+(4 \gamma -1)\upsilon(\alpha,\beta,\gamma)}{3 \alpha ^2 (7 \alpha ^2-2 \alpha  (\beta +2 \gamma -2)-\beta ^2)}
\,\\
y^{2}_{\pm}=\frac{(-4 \alpha ^3+\alpha ^2 (-2 \beta +2 \gamma -3)\pm\beta\upsilon(\alpha,\beta,\gamma)\alpha (\beta -2 \beta ^2-2 \beta  \gamma +\upsilon(\alpha,\beta,\gamma)))}{\alpha  \left(7 \alpha ^2-2 \alpha  (\beta +2 \gamma -2)-\beta ^2\right)}\,\\
z^{2}_{\mp}=\frac{\mp2\upsilon(\alpha,\beta,\gamma)-8 \alpha ^2+2 \alpha  (\beta +2 \gamma -3)+2 \beta ^2}{7 \alpha ^2-2 \alpha  (\beta +2 \gamma -2)-\beta ^2}
\end{cases}
\end{align}
where one has used
\begin{equation}
  \upsilon(\alpha,\beta,\gamma)=\sqrt{\alpha ^2 \left(2 \alpha ^2+\alpha  (-4 \beta -8 \gamma +2)+2 \beta ^2+\beta  (4 \gamma -2)+(1-2 \gamma )^2\right)}.
  \end{equation}
Concerning  the third non-BPS black hole   solution (\ref{3}), we get 
\begin{equation}
\begin{cases}
x^{3}_{\mp}=\frac{\mp F(\alpha,\beta,\gamma)}{\alpha ^2 \left(14 \alpha ^4-2 \alpha ^3 (18 \beta +36 \gamma +5)+\alpha ^2 \left(47 \beta ^2+12 \beta  (8 \gamma +1)+24 \gamma  (4 \gamma +1)\right)-6 \alpha  \beta ^2 (5 \beta +10 \gamma +1)+9 \beta ^4\right)}
\,\\
y^{3}_{\pm}=\frac{\pm G(\alpha,\beta,\gamma)}{\alpha  \left(14 \alpha ^4-2 \alpha ^3 (18 \beta +36 \gamma +5)+\alpha ^2 \left(47 \beta ^2+12 \beta  (8 \gamma +1)+24 \gamma  (4 \gamma +1)\right)-6 \alpha  \beta ^2 (5 \beta +10 \gamma +1)+9 \beta ^4\right)}
\,\\
z^{3}_{\pm}=-\frac{6 (-8 \alpha ^4+2 \alpha ^3 (5 \beta +10 \gamma -3)+\alpha ^2 (-2 \beta ^2+10 \beta +20 \gamma +2)\pm\sqrt{\alpha ^2 (-2 \alpha ^2-2 \alpha +\beta ^2)^2 \upsilon(\alpha,\beta,\gamma))}-\alpha  \beta ^2 (5 \beta +10 \gamma +7)+3 \beta ^4)}{14 \alpha ^4-2 \alpha ^3 (18 \beta +36 \gamma +5)+\alpha ^2(47 \beta ^2+12 \beta  (8 \gamma +1)+24 \gamma  (4 \gamma +1))-6 \alpha  \beta ^2 (5 \beta +10 \gamma +1)+9 \beta ^4}
\end{cases}
\end{equation}
where one has used 
{\small \begin{align*}
F(\alpha,\beta,\gamma)&=-28 \alpha ^6+\alpha ^5 (24 \beta +48 \gamma -46)+6 \alpha ^4(7 \beta ^2+\beta  (8 \gamma +13)+8 \gamma ^2+26 \gamma +2)\\
&-\alpha ^3(84 \beta ^3+\beta ^2 (168 \gamma +59)+6 \beta  (8 \gamma -1)+6 (8 \gamma ^2-2 \gamma -1))+\alpha ^2 \beta ^2(58 \beta ^2\\
&+\beta  (72 \gamma -3)+72 \gamma ^2-6 \gamma -15)\mp 6 \beta ^2 \sqrt{\alpha ^2 (-2 \alpha ^2-2 \alpha +\beta ^2)^2 \upsilon(\alpha,\beta,\gamma))}\\
&+3 \alpha  (\sqrt{\alpha ^2 (-2 \alpha ^2-2 \alpha +\beta ^2)^2 \upsilon(\alpha,\beta,\gamma))}(2 \beta  +(4 \gamma +1))-4 \beta ^5+\beta ^4 (4-8 \gamma ))\\
G(\alpha,\beta,\gamma)&=24 \alpha ^5-2 \alpha ^4 (34 \beta +30 \gamma -9)+2 \alpha ^3 (36 \beta ^2+\beta  (66 \gamma -19)-30 \gamma -3)\\
&+\alpha ^2 \beta  (-38 \beta ^2+\beta  (39-66 \gamma )-96 \gamma ^2+36 \gamma +6)+3 \beta  \sqrt{\alpha ^2 (-2 \alpha ^2-2 \alpha +\beta ^2)^2 \upsilon)}\\
&\pm3 \alpha  (\sqrt{\alpha ^2 (-2 \alpha ^2-2 \alpha +\beta ^2)^2 \upsilon)}-2 \beta ^4+\beta ^3 (5-10 \gamma)).
\end{align*}}

\begin{figure}[h!]
\begin{center}
\includegraphics[scale=1]{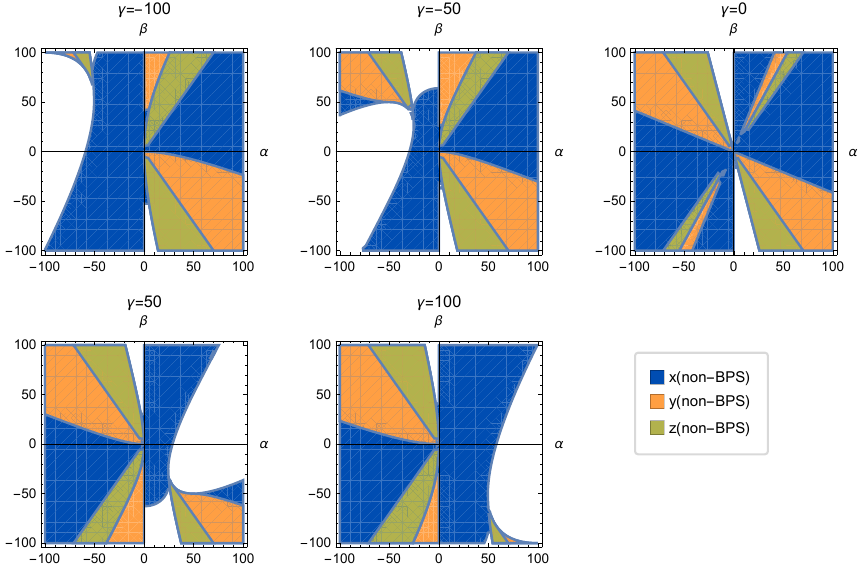}
\end{center}
\caption{{\it \footnotesize   Allowed electric charge regions for   the second non-BPS black hole solution for the triplet  $(x^2_-, y^2_-, z^2_- )$. }}
\label{F2}
\end{figure}

\begin{figure}[h!]
\begin{center}
\includegraphics[scale=1]{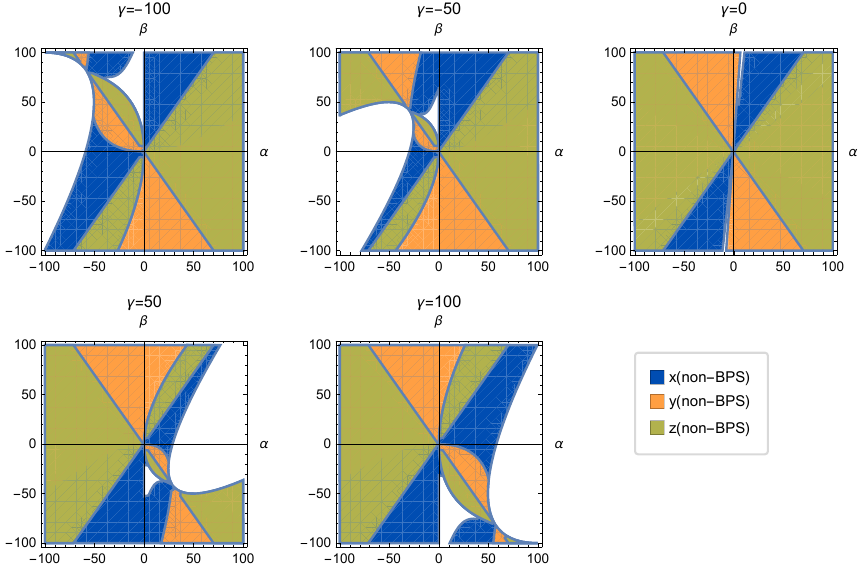}
\end{center}
\caption{{\it \footnotesize   Allowed electric charge regions for   the third non-BPS black hole solution for   the triplet $(x^3_-, y^3_-, z^3_- )$. }}
\label{F2f}
\end{figure}

In Fig.(\ref{F2}) and Fig.(\ref{F2f}), we illustrate  the allowed  electric charge regions in  the half cone configurations   for the second and third  solutions of the non-BPS black holes that are  constrained by  the K\"{a}hler  cone conditions, respectively. 

The two figures illustrate that the cone symmetry configuration is recovered when we set $\gamma = 0$, as observed for 5D   black holes obtained from M-theory on  CY threefold with two  K\"{a}hler  parameters \cite{1,4,6}. Non zero values  of  the  $\gamma$ breaks such a  symmetry  in the  graphical representation of  the allowed  regions of black hole electric  charges.  In fact, the positive and the negative charge values should be gathered to recover the symmetric configurations appearing in  M-theory on two parameter CY  manifolds.   Moreover, it has been remarked that  for  these two solutions,  however,  certain regions that are allowed for  the BPS black holes become disallowed.

Having discussed solutions  with  lower orders, we  move  now   to  approach the higher  ones associated with  the fourth and the  fifth solutions. Due to the higher order of the local geometric coordinates in the charge expressions, a direct study of the allowed charge regions  is not an easy task.  It may need  more   advanced numerical computations.  Nevertheless, it could be  possible to determine the range of the  charge ratios that admit attractor solutions lying within the  K\"{a}hler  cone.   A close examination shows that there are certain  singular curves  in the moduli space, which do not correspond to any possible extremal black hole configurations for  charge  finite values. These curves are obtained by setting the denominators in the expressions of $\alpha$, $\beta$, and $\gamma$  to zero. To unveil  such behaviors,  we consider  the fourth solution given in  (\ref{4}). Indeed,  it is evident that there is a common singularity among the three charges at $x = 2$. This contrasts with the previous study, where the singularity  has been  located within a region of the $(\alpha, \beta, \gamma)$ space. However, it is reduced  here to a single point in  such a  space. 

Moreover, the denominator of $\gamma$ in this form reveals the presence of  the complex singularities. A closer inspection indicates that additional singularities can be uncovered by performing  certain changes of variables, resembling coordinate singularities. From the solution (\ref{4}), we can  obtain the expressions of   $x$ and $z$ as follows
\begin{equation}
x = \frac{2 \alpha -2 \beta +y+1}{\alpha -\beta }, \qquad 
z = \frac{2 (\beta +\alpha  y)}{\alpha -\beta}.
\end{equation}
Replacing   $ x $ and $ z $ in the $ \gamma $ expression, we get
\begin{equation}
\label{6}
\gamma=\frac{s(y)}{\ell(y)}
\end{equation}
where $s(y)$ and $\ell(y)$  are found to be 
{\small \begin{align*}
\ell(y) &= 4 ( \beta +\alpha  y)(2 \alpha ^4 (6 y^4+22 y^3+35 y^2+24 y+6)+2 \alpha ^3(11 y^4+2 (\beta +23) y^3+(71-4 \beta ) y^2\\
&-2 ( \beta -24) y+12 )-\alpha ^2(2(7 \beta ^2-6)+(11 \beta ^2-12) y^4+(46 \beta ^2+4 \beta -48) y^3+(61 \beta ^2+8 \beta -72) y^2\\
&+4 (13 \beta^2+\beta -12) y)+2 \alpha  \beta ^2(-6 y^4+(\beta -24) y^3+(2 \beta -37) y^2+(5 \beta -26) y-7)\\
&+\beta ^4 (3 y^4+12 y^3+19 y^2+14 y+6))\\  \,\\
s(y)&=2 \alpha ^6 (35 y^5+99 y^4+92 y^3-36 y-12)+\alpha ^5 (48 \beta +(134-24 \beta ) y^5+(64 \beta +382) y^4+4 (71 \beta\\
& +69) y^3+24 (19 \beta -6) y^2+24 (11 \beta -10) y-72)+\alpha ^4 (12 (5 \beta ^2+10 \beta -6)+(-55 \beta ^2-44 \beta+68) y^5\\
&+(-179 \beta ^2+104 \beta +160) y^4+(-48 \beta ^2+692 \beta -4) y^3+12 (6 \beta ^2+89 \beta -24) y^2+12(13 \beta ^2+52\\ 
&\beta -22) y)+2 \alpha ^3 (-28 \beta ^3+60 \beta ^2+48 \beta +\beta(11 \beta ^2-23 \beta -12) y^5-2(9 \beta ^3+26 \beta ^2-21\beta +6) y^4\\
&+(-169 \beta ^3+33 \beta ^2+252 \beta -48) y^3-2 (95 \beta ^3-78 \beta ^2-189 \beta +36) y^2-6 (23 \beta ^3-29 \beta ^2-38 \beta +8) y\\
&-12)+2 \alpha ^2 \beta  (-20 \beta ^3-42\beta ^2+30 \beta +3 \beta(\beta ^2+4 \beta +2) y^5+3(2 \beta ^3-11 \beta ^2+10 \beta +4) y^4+(-9 \beta ^3\\
&-204 \beta ^2+78 \beta +48) y^3+(-66 \beta ^3-287 \beta ^2+120 \beta +72) y^2+(-34 \beta ^3-190 \beta ^2+96 \beta +48) y\\
&+12)-2 \alpha  \beta ^3 (-12 \beta ^2+20 \beta +3 \beta  (\beta +2) y^5+(-6 \beta ^2+21 \beta +12) y^4+(-39 \beta ^2+45 \beta +48) y^3\\
&+(-58 \beta ^2+70 \beta +74) y^2+(-26 \beta ^2+50 \beta +52) y+14)
+\beta ^5 (4 (\beta +3)+3 \beta  y^5+(9 \beta +6) y^4\\
&+6 (3 \beta +4) y^3+(28 \beta +38) y^2+4 (6 \beta +7) y).
\end{align*}}
Solving the  constraint $ \ell(y) =0 $,  we obtain the following  new real pole 
\begin{equation}
 y=-\dfrac{\beta}{\alpha}.
\end{equation}
In this case, the singularity is located in the region  with opposite charges being  either   ($\beta > 0$ and $\alpha < 0$)  or  ($\beta < 0$ and $\alpha > 0$).
These regions can reduced by replacing  $y$ in   the $x$ and  the $z$ expressions. This yields to 
\begin{equation}
x=\frac{1+2\alpha }{\alpha },  \quad z=0.
\end{equation} 
Now, we  would like to  discuss the poles of the fifth solution (\ref{5}). Interestingly, the singularity at $x = 2$ is  still  present here, appearing as a trivial pole. However,  non-trivial poles can be obtained by considering the  surface constraint  $\chi=0$.
Resolving this equation, we   can express  the poles of $ y $ as  functions of $ x $ and $ z $ as follows
\begin{equation}
\begin{cases}
y_{(\pm,\mp)}=\frac{1}{6} \left(3 z(\pm,\mp)\sqrt{3} \sqrt{(\pm,\mp)\iota(x,y,z)+6 x (z+2)+4 z^2-12}\right)\\
y_{(\mp,\mp)}=\frac{1}{6} \left(3 z(\mp,\mp)\sqrt{3} \sqrt{(\mp,\mp)\iota(x,y,z)+6 x (z+2)+4 z^2-12}\right),
\end{cases}
\end{equation}
where $\iota(x,y,z)= \sqrt{3} \sqrt{(z+2) \left(12 x^2 (z+2)+16 x \left(z^2-3\right)+5 z^3-10 z^2-12 z+24\right)} $.  It has been observed that there are four solutions $ (y_{(+-)},y_{(-+)}) $ and $ ( y_{(--)},  y_{(++)}) $. These poles satisfy the Kahler cone condition subject to the following constrain regions 
\begin{equation}
(x < 1 \,\,\,\, \text{or}   \,\,\,\,z \geqslant 2 - 2 x) \,\,\,\,    \text{and} \,\,\,\, (x \geqslant 1  \,\,\,\, \text{or} \,\,\,\,  z> 0).
\end{equation}
\subsection{Recombination factor and stability  of  non-BPS black holes}
\begin{figure}[h!]
\begin{center}
\begin{tabular}{cc}
\hline 
\includegraphics[scale=0.75]{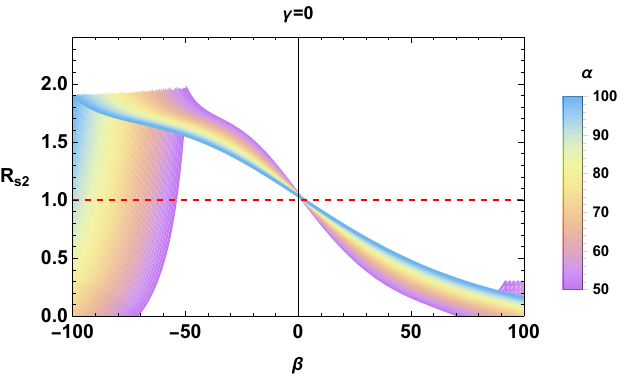}
& 
\includegraphics[scale=0.75]{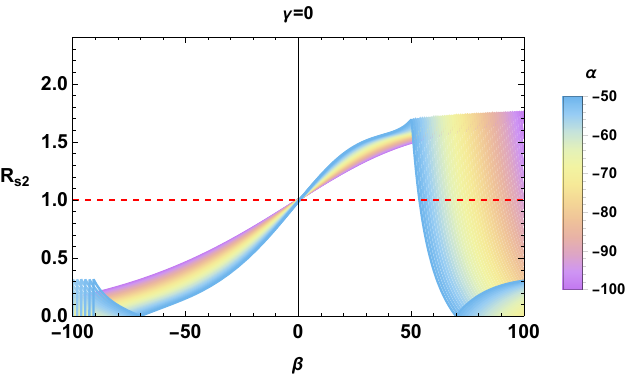}
\\ 
\hline 
\end{tabular} 
\caption{{{\it \footnotesize Recombination factor  for the second  solution in the allowed charge regions. }}}
\label{S1x}
\end{center}
\end{figure}

Having determined the  non-BPS black hole solutions, we move now to inspect their stability behaviors. This  analysis  can be done by calculating  the recombination factor $R$  via    a generic computation. Following to \cite{1}, this  factor of  such  non-BPS solutions      can be  determined   by means of  the  following relation
 \begin{equation}
 R=\frac{\sqrt{{V^{cr}_{eff}}}}{M_{C^{\cup}}}
 \end{equation}
where  $V^{cr}_{eff}$  denotes   the critical value of the  effective  potential $ V_{eff}$ and $M_{C^{\cup}}=\sum|\varrho_{I}|t_{I}$. To compute  $\varrho_{I}$, we consider 
the  non-BPS black holes obtained  by  wrapping  M2-branes on     a  non-holomorphic  curve class embedded in the proposed   toric CY threefold given by 
\begin{equation}
C=\varrho_{1}C^{1}+\varrho_{2}C^{2}+\varrho_{3}C^{3}+\varrho_{4}C^{4}. 
\end{equation}
The associated   black hole charges  are  given by 
\begin{equation}
 q_{I}=\int_{C}J_{I}= \varrho_{1}C^{1}J_{I}+\varrho_{2}C^{2}J_{I}+\varrho_{3}C^{3}J_{I}+\varrho_{4}C^{4}J_{I}, \qquad I=1,\ldots,4. 
  \end{equation} 
Using  the  K\"{a}hler cone matrix, we get 
\begin{eqnarray}
   \varrho_{1}&=&q_{4}-2(q_{1}+q_{2})\nonumber \\
   \varrho_{2}&=& -q_{1}\\
   \varrho_{3}&= &q_{3}\nonumber \\
   \varrho_{4}&= &q_{1}+q_{4}.  \nonumber
 \end{eqnarray}
After appropriate computations, the recombination factor  is found to be 
\begin{equation}
R=\frac{1}{\sqrt{2} (| \alpha +\beta | +x | 1-2 (\alpha +\beta )| +y | \alpha | +z | \gamma | )}\sqrt{\frac{3\zeta}{x (z+2)-y^2+y z-2}}
\end{equation} 
where one has used
{\small 
\begin{align*}
\zeta&=-24 \alpha ^2-24 \alpha +12 \beta ^2+6 \alpha ^2 x^3 (z+2)+x^2(\alpha ^2(-6 y^2+6 y z+7 z^2-36)+6 \alpha  \beta \\
& (z+2) (2 y-z)+3 \beta ^2 (z+2)^2)+x (12(-2 \beta ^2+2 \alpha ^2(y^2+2)-\alpha (y^2+2) (\beta  y-1)+2 \beta  y+1)\\
&-6 z (2 \beta ^2-\alpha (2 \beta +3 \beta  y^2+y^2+4 \beta  y-2 y+6)+4 \alpha ^2 y+\beta  (2-4 y)-3)+2 z^2 (10 \alpha ^2+4\\
& \beta ^2+6 (\gamma ^2-\gamma +1)+\alpha  (-12 \beta -12 \gamma +(4 \beta -3) y+17)+3 \beta  (2 \gamma +y-3))+z^3 (14 \alpha ^2\\
&-2 \alpha  (7 \beta +7 \gamma -6)+4 \beta ^2+6 \beta(\gamma -1)+6 \gamma ^2-6 \gamma +3))-3 y^4 (2 \alpha ^2+2 \alpha +\beta ^2)\\
&+6 y^3(z(2 \alpha ^2-2 \alpha  (\beta -1)+(\beta -1) \beta)-2 \beta)-y^2(6 (2 \alpha +1)^2+z^2(16 \alpha ^2-4 \alpha  (6 \beta\\
& +3 \gamma -4)+6 \beta ^2+6 \beta  (\gamma -2)+6 \gamma ^2-6 \gamma +3)+6 z (2 \alpha -3 \beta +1))+y(-24 \beta+z^3(10 \alpha ^2\\
&-2 \alpha (4 \beta +6 \gamma -5)+2 \beta ^2+4 \beta  (\gamma -1)+6 \gamma ^2-6 \gamma +3)+2 z^2 (6 \alpha -2 \beta +3)+6 (2 \alpha +1)\\
& z (2 \alpha -2 \beta +1))+4 \alpha ^2 z^4-4 \alpha  \beta  z^4-4 \alpha  \gamma  z^4+4 \alpha  z^4+\beta ^2 z^4+2 \beta  \gamma  z^4-2 \beta  z^4+\gamma ^2 z^4\\
&-2 \gamma  z^4+z^4+4 \alpha  z^3-2 \beta  z^3-2 \gamma  z^3+2 z^3-20 \alpha ^2 z^2+24 \alpha  \beta  z^2+24 \alpha  \gamma  z^2\\
&-20 \alpha  z^2-8 \beta ^2 z^2-12 \beta  \gamma  z^2+12 \beta  z^2-12 \gamma ^2 z^2+12 \gamma  z^2-5 z^2-24 \alpha  z+12 \beta  z-12 z-12.
\end{align*}}
For simplicity reasons, we   analyse the  stability behavior within the allowed electric  charge regions of   the second and  the third non-BPS solutions presented previously. Considering the second solution  (\ref{2}), we illustrate in Fig.(\ref{S1x}) the behavior of the recombination factor  by taking  $\gamma = 0$ and  varying  $\beta$ between $-100$ and $100$ for positive and negative  $\alpha$ charges given in the left and right panels, respectively.

 To perform this study in the allowed regions, we select appropriate values of $\alpha$ providing two graphs.   The latters  exhibit  a remarkable  symmetry  with one appearing as the mirror image of the other. This  can be supported by  the  cone symmetry  observed in the allowed charge region illustrations. 
 It follows that the recombination factor R rises to its maximum values in the left and  right plots in the respective regions $-100<\beta <-50 $ and $50<\beta <100 $  where   the stable and unstable black holes  live.   Taking  $-50<\beta <0 $ and $0<\beta <50 $,  we observe that  the black holes are unstable,  for positive and negative $\alpha$ values, respectively. The cited symmetry  can be understood from  the fact that the regions with stable black holes  are associated  the  positivity and  the negativity behaviors  of  $\beta $ in the left  and the right plots,  respectively. In addition, for $\gamma =  \beta = 0$, a critical point appears in both graphs where all the curves meet. Interestingly, at this point, the behavior of the black hole changes from stable to unstable, and vice versa.

\begin{figure}[h!]
\begin{center}
\begin{tabular}{cc}

\hline 
\includegraphics[scale=0.75]{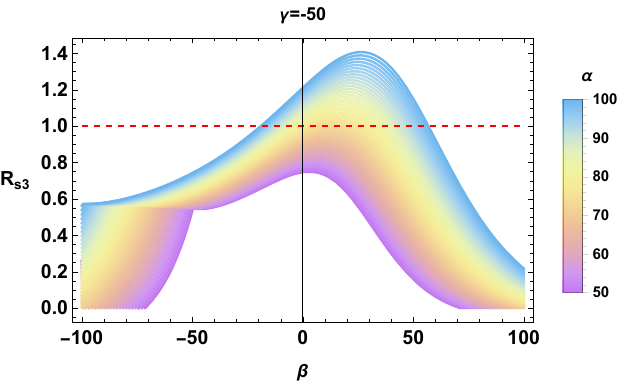}
& 
\includegraphics[scale=0.75]{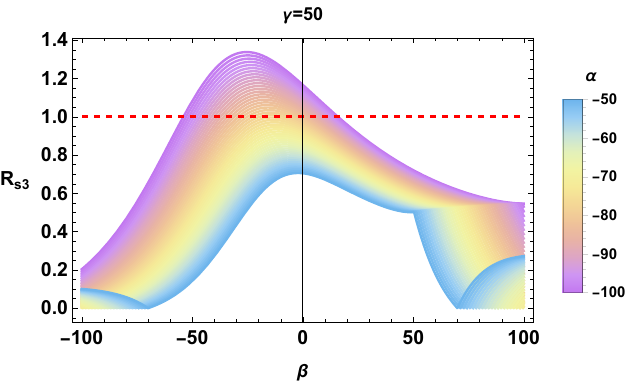}
\\ 
\hline 
\end{tabular} 
\caption{{{\it \footnotesize Recombination factor  for the   third solution in the allowed charge regions. }}}
\label{S21x}
\end{center}
\end{figure}

For the third solution  (\ref{3}),  we  approach this symmetrical behavior by taking $\gamma \neq 0$. This means that for each negative value of $\gamma$, we can associate a symmetrical positive value with the same magnitude and  then we study the  recombination factor  behaviors.  Effectively, we consider $\gamma = 50$ and $\gamma = -50$. In Fig(\ref{S21x}), we  illustrate the variation of the recombination factor for opposite charge regions. 

As the figure shows, the same symmetrical behavior is present, but not for the same $\gamma$ value. This symmetry is observed for each pair of opposite values, which could be linked to the symmetry observed in the allowed region where, each half-cone configuration is associated with the second half-cone configuration that has  the opposite charges, recovering the full cone configuration.  Since it has been shown  that a solution could be stable if this factor is more than one,   this  figure provides  stable and unstable black hole configurations depending on the allowed charge regions. Indeed,  for $ R < 1$,  the constituent BPS-anti-BPS pairs  can  recombine to  generate  stable non-BPS black hole states.

\section{M-theory black strings from  a  THCY with   $ h^{1,1}=4$}
In this section, we   study 5D   black strings using   the  M-theory compactification on    the proposed  THCY with   $ h^{1,1}=4$.  These solutions  can be provided by    wrapping  M5-branes on  dual divisors   of such a  toric CY  manifold.    The associated stability behaviors can be discussed   by means  of  the  effective  potential given in terms of the M5-brane  magnetic  charges.  
\subsection{5D black string solutions}
Before approaching the stability behaviors,   one needs first to   find  the black string solutions carrying  four  magnetic charges $p^I$,   $I=1,2,3,4$ under $U(1)^4$ gauge symmetries.   To do so,   we should   calculate
the  effective potential  $V_{eff}^{m} $ of  the 5D black strings. Indeed, it is given by 
\begin{equation}
V_{eff}^{m}=\frac{J}{t_3^2 (-3 t_2^2+3 t_3 t_2+t_3^2-6 t_4^2+3 t_1 (t_3+2 t_4))^2}
\end{equation}
where the scalar function  $J$  is expressed as follows 
\begin{align*}
J=&6 (p_3^2(3 t_2^4-6 t_3 t_2^3+6(t_3^2+2 t_4^2-t_1 (t_3+2 t_4)) t_2^2+4 t_3((t_3^2-3 t_4^2+3 t_1 ((t_3+t_4)) t_2+t_3^4+12 t_4^4\\
&+6 t_1^2(t_3^2+2 t_4 t_3+2 t_4^2)+4 t_1 (t_3^3-3 t_4^2 t_3-6 t_4^3))+2 p_3 t_3^2(2 p_4 (3 t_1+3 t_2+2 t_3)(t_1-2 t_4)\\
&+p_1(3 t_2^2+6 t_4 t_2+t_3^2+6 t_4^2+4 t_3 t_4)+p_2(-3 t_2^2-4 t_3 t_2+t_3^2+6 t_4^2-6 t_1(t_2+t_4)))\\
&+t_3^2 (p_2^2(6 t_2^2-6 t_3 t_2+5 t_3^2-12 t_4^2+6 t_1(t_3+2 t_4))-6 p_1 p_2(2 t_2-t_3)(t_3+2 t_4)+3 p_1^2(t_3+2 t_4)^2
\\
&+4 p_4^2(3 t_1^2+3(t_3-2 t_4) t_1-3 t_2^2+t_3^2+6 t_4^2+3 t_2 t_3)-4 p_4(3 p_2(2 t_2-t_3) (t_1-2 t_4)\\
&+p_1(-3 t_2^2+3 t_3 t_2+t_3^2+6 t_4^2+6 t_3 t_4))))).
\end{align*}
Now,  we need  to determine the    critical points in terms of   the local  coordinate of the moduli space.   Roughly, one has the following new  dynamical  variables 
\begin{equation}
 \rho=\frac{p_{1}}{p_{4}}, \qquad  \sigma=\frac{p_{2}}{p_{4}}, \qquad  \tau=\frac{p_{3}}{p_{4}} \qquad  x=\frac{t_1}{t_4}, \qquad   y=\frac{t_{2}}{t_4},  \qquad  z=\frac{t_3}{t_4}.
\end{equation}
The black string solutions can  be  approached by considering  $ t^{I}=\frac{3p^{I}}{Z_{m}}$, where $ Z_{m} $ represents the central charge of the  black  strings. In this way, the BPS solutions  are given in terms of  the magnetic charges 
\begin{equation}
 x=\rho, \qquad  y=\sigma, \qquad  z=\tau.
 \label{sBS1}
\end{equation}
The above solutions show that all the BPS region charges inside  the  K\"{a}hler cone corresponding  to the positivity of  the geometric  local variables $ x $, $ y $ and $ z $.

 To  establish  the  equations of motion  of   the scalar fields  associated with the black string moduli space,  one needs to  employ
 an   extremization  mechanism with respect  to the  geometric local  variables  via  (\ref{defVeff}). Indeed, we find three equations which  can be  formulated  as follows
 {\small  \begin{align}
24 z (2 \sigma -\tau -(\tau +2) y+\sigma  z+z) (3 (\tau +2) x-6 \sigma  y+3 \tau  y+3 \rho  (z+2)+3 \sigma  z+2 \tau  z-12)=0  \\
-24 z (2 (\rho +\tau )-(\tau +2) x+(\rho -2) z) (3 (\tau +2) x-6 \sigma  y+3 \tau  y+3 \rho  (z+2)+3 \sigma  z+2 \tau  z-12)=0
\end{align}}
and
 {\small 
 \begin{align}
&-24 (z^3(10 (\rho -2) (3 \rho +3 \sigma +2 \tau )-3 x^2 (\sigma ^2+2 \sigma  (\tau +3)+\tau ^2+8 \tau +10)+x (-12 \sigma  (\rho -\tau -5)\nonumber\\
&+4 \rho  \tau +6 \sigma ^2+6 \tau ^2+40 \tau +3 y (2 \rho  (\sigma +\tau +3)+5 \sigma ^2+2 \sigma  (\tau +4)-\tau ^2-8 \tau -10)+60)\nonumber\\
&-(\rho -2) y^2 (3 \rho +15 \sigma -4 \tau )+6 y (2 \rho ^2-2 \rho  (3 \sigma -\tau +5)-5 \sigma ^2-2 \sigma  \tau +\tau ^2+10))+6 \tau ^2 (x-2) z \nonumber\\
&(2 x^2-x (y^2-2 y+2)-y (y^2+2))-3 z^2 (-4 (\rho ^2+\rho  (\tau -2)-\sigma ^2+\sigma  \tau +\tau ^2+2)+4 (\tau +1) x^3\nonumber\\
&-2 x^2 (-\sigma ^2+\sigma  \tau +\tau ^2+6 \tau +2 \sigma  (\tau +2) y-2 \tau  y+6)+x (6 (-\sigma ^2+\sigma  \tau +\tau ^2+2 \tau +2)\nonumber\\
&+y^2 (2 \rho  (\tau +2)+3 \sigma ^2-3 \sigma  \tau +\tau ^2-2 \tau -6)+y (-4 \rho  \sigma +2 (\rho -6) \tau +8 \sigma  (\tau +3)))-(\rho -2) y^3\nonumber \\
&(2 \sigma -\tau )+2 y^2 (\rho ^2-\rho  (\tau +6)-3 \sigma ^2+3 \sigma  \tau -\tau ^2+6)+6 (\rho -2) y (2 \sigma -\tau ))+3 \tau ^2 (x-2)\nonumber\\
& (-2 x+y^2+2)^2-z^4 (x (\rho  (6 \sigma +4)+7 \sigma ^2+\sigma  (7 \tau +12)+2 (\tau ^2+7 \tau +7))-6 \rho ^2 (y+3)+\rho  (-24 \sigma\nonumber\\
& -14 \tau +(12-7 \tau ) y+28)-2 (7 \sigma ^2+7 \sigma  \tau +2 (\tau ^2+7)-7 \tau  y))+(\rho -2) z^5 (3 \rho +3 \sigma +2 \tau ))=0.\nonumber\\
 \end{align}}
 Solving these scalar  equations of motion, we  can obtain the magnetic  black string configurations in terms the geometric ones. They are organized as follows:\\
Solution 1
\begin{eqnarray}
\label{Fbs2}
 \rho &=&\frac{\omega(x,y,z)}{(z+2) \left(12 x \left(z^2+3 z+2\right)-12 y^2 (z+1)+12 y (z+1) z+4 z^3+5 z^2-24 z-24\right)} \nonumber\\
 \sigma &=&\frac{y \left(12 x (z+2)-7 z^2-24\right)-4 z \left(3 x (z+2)+z^2-6\right)-12 y^3+24 y^2 z}{12 x \left(z^2+3 z+2\right)-12 y^2 (z+1)+12 y (z+1) z+4 z^3+5 z^2-24 z-24} \\
 \tau&=&-\frac{3 z \left(4 x (z+2)-4 y^2+4 y z+z^2-8\right)}{12 x \left(z^2+3 z+2\right)-12 y^2 (z+1)+12 y (z+1) z+4 z^3+5 z^2-24 z-24}\nonumber
\end{eqnarray}
where one has used 
{\small 
\begin{align*}
\omega(x,y,z)&=-12 x^2 (z+2)^2+3 x (z+2) \left(12 y^2-12 y z+5 z^2+16 z+24\right)-24 y^4+48 y^3 z\\
&-2 y^2 \left(17 z^2+24 z+48\right)+2 y z \left(5 z^2+24 z+48\right)+6 z^4+16 z^3-20 z^2-96 z-96.
\end{align*}}
Solution 2
\begin{eqnarray}
\label{Fbs1}
\rho &= &\frac{-3 x (z+2)+6 y^2-6 y z-2 z^2+12}{3 (z+2)}\nonumber \\
\sigma &=&y \\
 \tau &= &z. \nonumber
\end{eqnarray}
Solution 3
{\small 
\begin{eqnarray}
\label{Fbs3}
 \rho &=& \frac{\Phi}{\Omega}\nonumber\\
 \sigma &=&\frac{\Psi}{\Omega} \\
 \tau&=&\frac{z \left(-3 x (z+2)+3 y^2-3 y z+z^2+6\right) \left(4 x (z+2)-4 y^2+4 y z+z^2-8\right)}{\Omega} \nonumber
\end{eqnarray}}
where one has used
{\small 
\begin{align*}
\begin{cases}
\Phi =12 x^3 (z+2)^2+3 x^2 (z+2)(-8 y^2+8 y z+11 z^2+16 z-16)+x (12 y^4-24 y^3 z\\
+y^2(-45 z^2-96 z+48)+3 y z (19 z^2+32 z-16)+9 z^4+16 z^3-114 z^2-192 z+48)\\
+8 z (3 y^4-6 y^3 z+2 y^2 (z^2+6)+y z (z^2-12)-2 (z^2-6))\\
  \,\\
\Psi =3 y(4 x^2 (z+2)^2-x (5 z^3+10 z^2+16 z+32)-z^4+10 z^2+16)-4 z (3 x^2 (z+2)^2\\
+x (z^3+2 z^2-12 z-24)-2 (z^2-6))-3 y^3 (8 x (z+2)-9 z^2-16)+y^2 z (48 x (z+2)+z^2-96)\\+12 y^5-36 y^4 z \\
  \,\\
 \Omega=12 x^2 (z+1) (z+2)^2+x (z+2)(-24 y^2 (z+1)+24 y (z+1) z+4 z^3+9 z^2-48 z-48)\\
+12 y^4 (z+1)-24 y^3 z (z+1)+y^2(8 z^3+3 z^2+48 z+48)+y z (4 z^3+9 z^2-48 z-48)\\+z^4-8 z^3-18 z^2+48 z+48.
\end{cases}
\end{align*}}
Solution 4
{\small  \begin{eqnarray}
\label{Fbs4}
 \rho &= &\frac{\Gamma_{\pm}}{(2 + z)\Lambda}\nonumber\\
 \sigma &=&-\frac{\Pi_{\pm}}{(2 y-z)\Lambda} \\
 \tau&=&3\frac{ \mp\sqrt{2} \eta -2 z^3 (6 x^2+6 x y+x-3 y^2-6 y-7)-12 (x-1) z^2 (2 x-y^2-2)+z^4 (-10 x-6 y+3)-2 z^5}{\Lambda} \nonumber
\end{eqnarray}}
where one has used 
{\small 
\begin{align*}
\begin{cases}
\Gamma_{\pm} =6 x^2 (z+2) (-6 y^2 (z+2)+6 y z (z+2)+z^2 (5 z+9))-x (\pm 3 \sqrt{2} \eta -36 y^4 (z+2)+72 y^3 (z+2) z\\
+6 y^2 (7 z^3+26 z^2+12 z-24)-6 y (13 z^3+38 z^2+12 z-24) z-22 z^5-29 z^4+156 z^3+252 z^2)\\
+2 (\pm 3 \sqrt{2} \eta +18 y^4 (z+2) z-36 y^3 (z+2) z^2+6 y^2 (z^3+5 z^2+12 z+12) z\\+6 y (2 z^3+z^2-12 z-12) z^2+2 z^6-2 z^5-24 z^4+12 z^3+72 z^2)  \\
  \,\\
\Pi_{\pm} = -6 y^2 (12 x^2 (z+2)^2+4 x (z^3-2 z^2-12 z-24)-z^4-8 z^3+8 z^2+48)-3 y\\ (\pm\sqrt{2} \eta -6 (2 x^2+5 x-3) z^3+(-72 x^2+48 x+24) z^2+(1-10 x) z^4-96 (x-1)^2 z-2 z^5)\\
-2 (18 x^2 z^3+36 x^2 z^2+15 x z^4-72 x z^2\pm\sqrt{2} \eta  (3 x+z-3)+3 z^5-6 z^4-18 z^3+36 z^2)\\+36 y^4 (z+2) (2 x+z-2)-18 y^3 z (3 z+8) (2 x+z-2) \\
  \,\\
 \Lambda= z^2 (18 x^2 (z+2)+6 x(2 z^2-2 z-15)+2 z^3-8 z^2-15 z+54)+6 y^2(-6 x (z^2+2 z+2)\\
 +z^3+8 z^2+6 z+12)+12 y z(3 x (z^2+2 z+2)+z^3-z^2-3 z-6)+18 y^4 (z+2)-36 y^3 z (z+2)
 \\  \,\\
 \eta=\sqrt{z^2 (z+2) (2 x+z-2) (z-2 y)^2 (3 x (z+2)-3 y^2+3 y z+z^2-6)^2}. 
\end{cases}
\end{align*}}

\begin{figure}[h!]
\begin{center}
\includegraphics[scale=1]{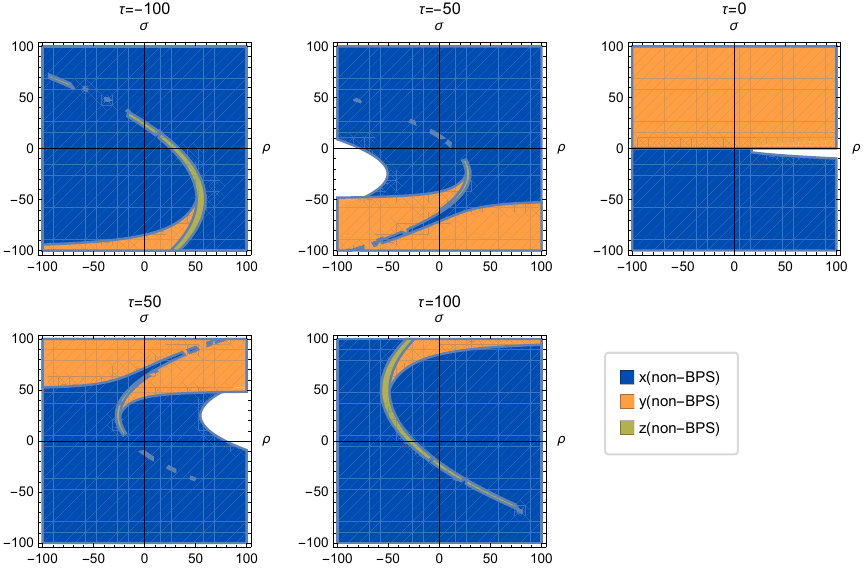}
\end{center}
\caption{{\it \footnotesize   Allowed magnetic charge regions for   the first non-BPS black string solution by considering $x, y$ and $z$  functions.}}
\label{FS1}
\end{figure}

As  in the previous cases, we should  express the local geometric
variable $x$, $y$ and $z$  in terms of the obtained  magnetic ratio charges of the 5D black strings.   This is needed   to  determine the allowed magnetic charge regions.
As shown in the above  equations above,  it has been inspected that  only the first and  the second solutions can  provide explicit expressions for the triplet \((x, y, z)\) in terms of the magnetic charge regions. They are given respectively by 
\begin{eqnarray}
\label{bss1}
x =\frac{-3 \rho  (\tau +2)+6 \sigma ^2-6 \sigma  \tau -2 \tau ^2+12}{3 (\tau +2)},\,\,\,y=\sigma ,\,\,\,z=\tau
\end{eqnarray}
and 
 {\small 
\begin{eqnarray}
\label{bss2}
x &=&\frac{B}{(\tau +2) \left(12 \rho  \left(\tau ^2+3 \tau +2\right)-12 \sigma ^2 (\tau +1)+12 \sigma  (\tau +1) \tau +4 \tau ^3+5 \tau ^2-24 \tau -24\right)}\nonumber\\
y &=&\frac{\sigma  \left(12 \rho  (\tau +2)-7 \tau ^2-24\right)-4 \tau  \left(3 \rho  (\tau +2)+\tau ^2-6\right)-12 \sigma ^3+24 \sigma ^2 \tau }{12 \rho  \left(\tau ^2+3 \tau +2\right)-12 \sigma ^2 (\tau +1)+12 \sigma  (\tau +1) \tau +4 \tau ^3+5 \tau ^2-24 \tau -24} \\
z&=& -\frac{3 \tau  \left(4 \rho  (\tau +2)-4 \sigma ^2+4 \sigma  \tau +\tau ^2-8\right)}{12 \rho  \left(\tau ^2+3 \tau +2\right)-12 \sigma ^2 (\tau +1)+12 \sigma  (\tau +1) \tau +4 \tau ^3+5 \tau ^2-24 \tau -24}\nonumber
\end{eqnarray}}
where one has used {\small 
\begin{align*}
B&=-12 \rho ^2 (\tau +2)^2+3 \rho  (\tau +2)(12 \sigma ^2-12 \sigma  \tau +5 \tau ^2+16 \tau +24)-24 \sigma ^4+48 \sigma ^3 \tau -2 \sigma ^2(17 \tau ^2\\
&+24 \tau +48)+2 \sigma  \tau  (5 \tau ^2+24 \tau +48)+6 \tau ^4+16 \tau ^3-20 \tau ^2-96 \tau -96(5 z^2+24 z+48)+6 z^4\\
&+16 z^3-20 z^2-96 z-96.
\end{align*}}
Using  the  black hole similar  techniques, in Fig.~(\ref{FS1}), we illustrate the allowed magnetic charge regions for the first solution given in  (\ref{bss1}).

As for black holes, we   observe  overlapping regions with symmetric K\"{a}hler cones.  They are given by three colors associated with the three  local geometric variables $x$, $y$ and $z$.

\begin{figure}[h!]
\begin{center}
\includegraphics[scale=1]{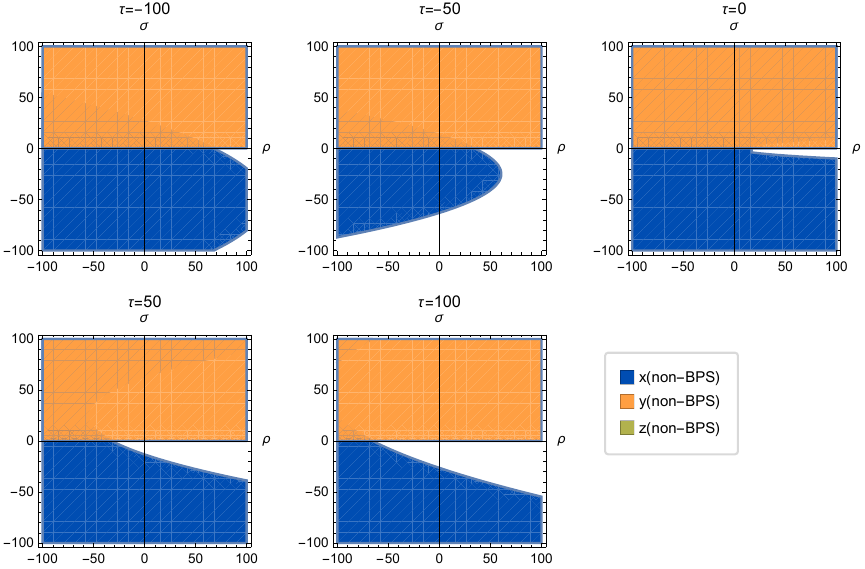}
\end{center}
\caption{{\it \footnotesize   Allowed magnetic charge regions for   the second non-BPS black string solution by considering $x, y$ and $z$ functions.}}
\label{FS2}
\end{figure}

\begin{figure}[h!]
\begin{center}
\includegraphics[scale=0.6]{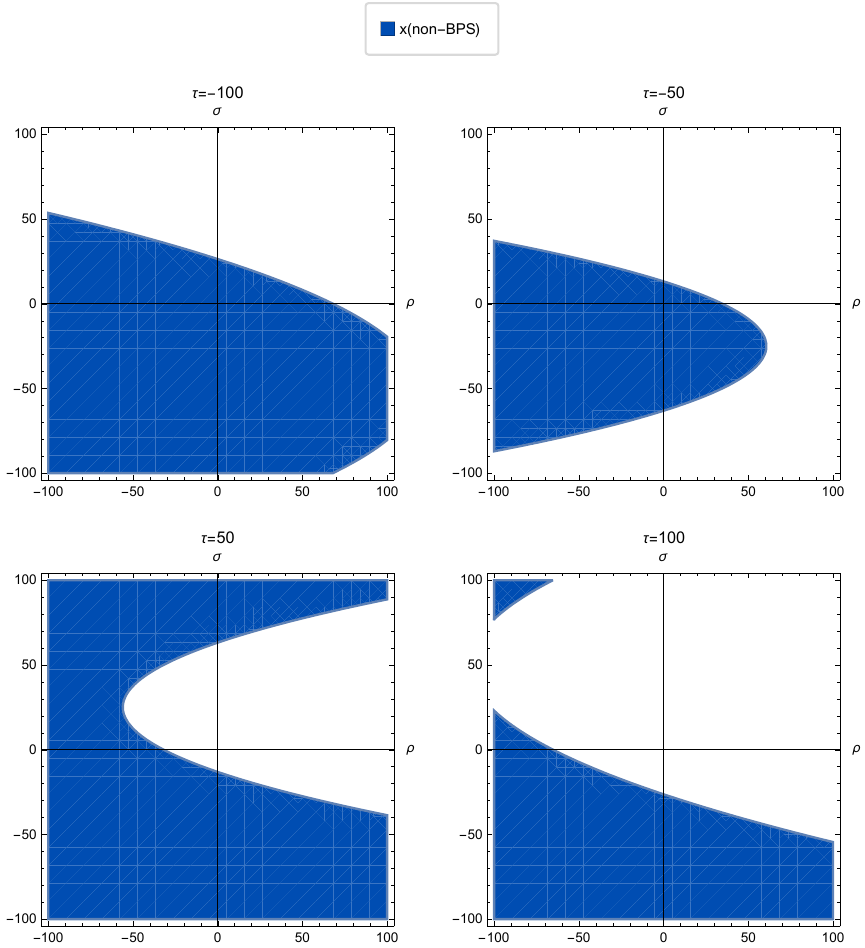}
\end{center}
\caption{{\it \footnotesize   Allowed magnetic charge regions for   the second non-BPS black string solution by considering only the $x$  variable.}}
\label{FS3}
\end{figure}

In Fig.~(\ref{FS2}), we present the allowed magnetic charge regions for the second solution
of non-BPS black strings given  in  (\ref{bss2}).  At first sight, it looks that the cone  symmetrical behavior  is broken.  A close observation, however,  shows that such a behavior still exists  by focusing 
  on  the  $x$ regions and the non-allowed charge regions. To make this more obvious, we illustrate in Fig.(\ref{FS3})  the allowed charge regions for the second solution by considering only  the local variable 
$x$.  It is clear from the figure that the cone symmetry behavior is recovered.

Before approaching the  stability behaviors,  we provide a   comment on   the non-BPS charges  associated with the equations (\ref{Fbs3}) and (\ref{Fbs4}). As we have done for the  black hole solutions, we consider  the pole behaviors obtained by resolving the vanishing  denominators $\Omega=0 $ and $ \Lambda=0$. They are found to be
\begin{equation}
\label{3}
\begin{cases}
 x_{\pm}^{p1} =\frac{24 y^2 (z+1)-24 y (z+1) z-4 z^3-\left(\pm\sqrt{16 z^2+24 z+33}+9\right) z^2+48 z+48}{24 (z+1)(z+2)}\\
  \,\\
  x_{\mp}^{p2}=\frac{6 y^2 \left(z^2+2 z+2\right)\mp\sqrt{3} \sqrt{(z-2 y)^2 \left(12 y^2 (z+1)^2-12 y (z+1)^2 z+\left(2 z^2+4 z+3\right) z^2\right)}-6 y \left(z^2+2 z+2\right) z-2 z^4+2 z^3+15 z^2}{6 z^2 (z+2)}.
\end{cases}
\end{equation}

\newpage

\subsection{Stability scenarios of  5D black strings}

To examine the stability of these solutions, we need to count the recombination factor, which is the ratio between the black string tension  $T$  and the minimum size of the corresponding piecewise calibrated divisors. By employing the local magnetic  variables, we can determine such a factor from the size  $ V_{D^{\cup}} $ which is the minimum piecewise calibrated volume representative of the class  $ [D] $  given by
 \begin{equation} 
 D=p_{1} {\cal D}_1+p_{2}{\cal D}_2 +p_{3}{\cal D}_3 +p_{4}{\cal D}_4
 \end{equation}
  where ${\cal D}_1$, ${\cal D}_2 $,  ${\cal D}_3$, ${\cal D}_4$ are dual divisors  to    $ {\cal J}_1$,  ${\cal J}_2$, ${\cal J}_3$   and  ${\cal J}_4$  denoting  the  K\"{a}hler  $(1,1)$-forms  of the proposed toric CY manifold,  respectively.   Roughly,  the   recombination  factor   is  expressed as  follows
 \begin{equation}
R=\frac{T}{ V_{D^{\cup}}}
\end{equation} 
where one has used  $ V_{D^{\cup}}=\sum\limits _{I=1}^4A_{I}|p_{I}|$ with  $ A_{I}=C_{IJK}t^{J}t^{K}=2\tau_{I} $ describing the size of the involved  divisors. Computations  lead to 
\begin{equation}
V_{C^{\cup}}= t_3( t_3+2t_4)|p_{1}|+t_3( -2t_2+t_3)|p_{2}|+ (2t_1( t_3+t_4)+t_3(2 t_2+t_4)- t_2^2-2 t_4^2) |p_{3}|+2t_3( t_1-2t_4)|p_{4}|.
\end{equation}
Combining these relations, the   black string  recombination factor   is found to be 
{\small{
\begin{equation}
R=\frac{\sqrt{\mathcal{F}}}{| \tau |  \left(2 x z+2 x-y^2+2 y z+z^2-2\right)+| \sigma |  \left(z^2-2 y z\right)+\left(z^2+2 z\right) | \rho | +2 x z-4 z}
\end{equation}}}
where one has used 
{\small 
\begin{align*}
\mathcal{F}(x,y,z)&=6 z^2 (2 (\rho ^2+\rho  (\tau -2)-\sigma ^2+\sigma  \tau +2)+(\tau ^2+2 \tau +2) x^2-2 x(-\sigma ^2+\sigma  \tau +2 \tau +\sigma  (\tau +2) y\\
&-\tau  (\tau +1) y+2)+y^2 (\rho  (\tau +2)+\sigma ^2-\sigma  \tau +\tau ^2-2)-2 (\rho -2) y (2 \sigma -\tau ))+6 \tau ^2 z (2 x^2\\
&-x (y^2-2 y+2)-y (y^2+2))+3 \tau ^2 (-2 x+y^2+2)^2+2 z^3 (2 (\rho -2) (3 \rho +3 \sigma +2 \tau )\\
&+x (3 \sigma ^2+6 \sigma +2 \tau ^2+4 \tau +6)-y (6 \rho  (\sigma +1)+3 \sigma ^2+4 \sigma  \tau -2 \tau ^2-6))+z^4(3 \rho ^2+2 \rho  (3 \sigma +\tau -2)\\
&+5 \sigma ^2+2 \sigma  \tau +\tau ^2+4).
\end{align*}}
\begin{figure}[h!]
\begin{center}
\begin{tabular}{cc}
\hline 
\includegraphics[scale=0.75]{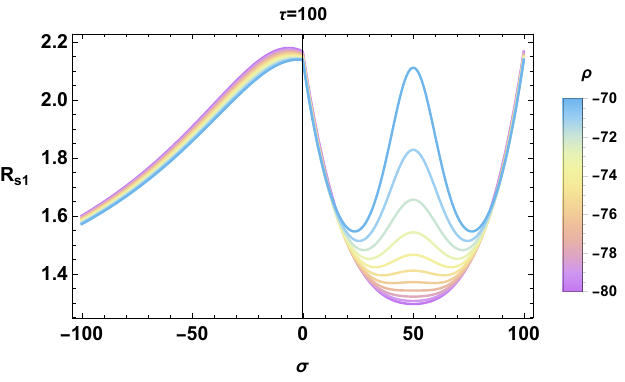}
& 
\includegraphics[scale=0.75]{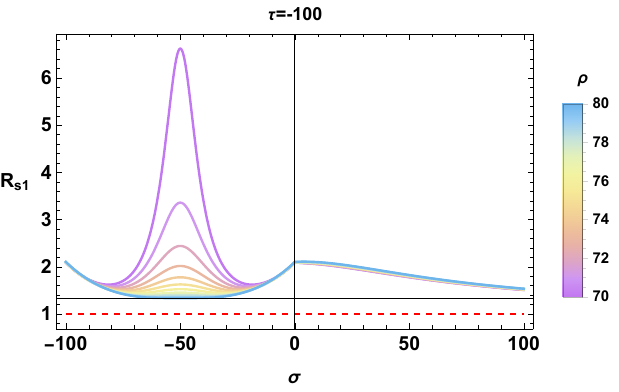}
\\ 
\hline 
\end{tabular} 
\caption{{{\it \footnotesize Recombination factor  for the first solution in the allowed magnetic  charge regions. }}}
\label{S11x}
\end{center}
\end{figure}

In Fig.(\ref{S11x}), we illustrate the behavior of the recombination factor in the allowed magnetic charge regions of the first black string solution by fixing the value of $\tau$ and varying $\rho$ and $\sigma$.
From the figure, it is evident that for different allowed regions of the magnetic charge, the black strings exhibit instability behaviors as the recombination factor exceeds 1. Moreover, a certain symmetry is observed in these two graphs, particularly for opposite values of $\tau$ and $\rho$ local magnetic charges.

\begin{figure}[h!]
\begin{center}
\begin{tabular}{cc}
\hline 
\includegraphics[scale=0.75]{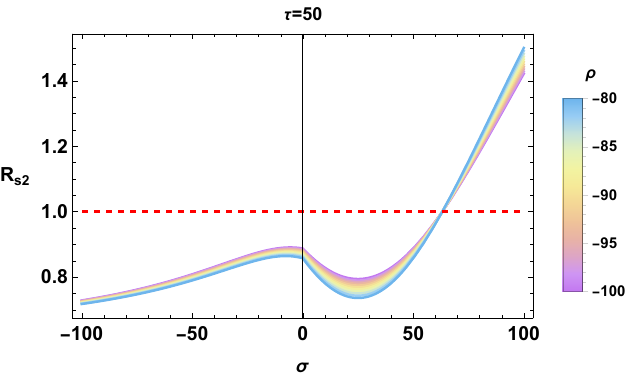}
& 
\includegraphics[scale=0.75]{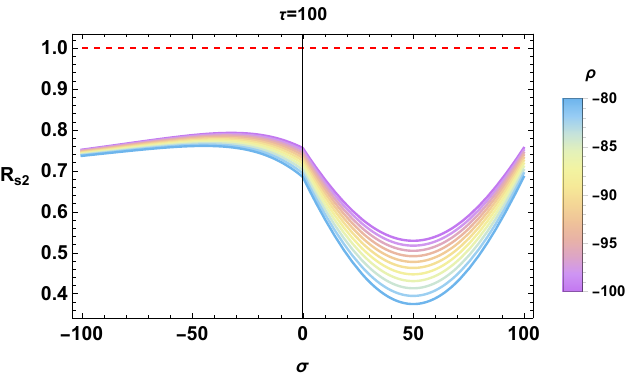}
\\ 
\hline 
\end{tabular} 
\caption{{{\it \footnotesize Recombination factor  for the second  solution in the allowed  magnetic charge regions. }}}
\label{S22x}
\end{center}
\end{figure}

Considering the second black string solution, the variation of the recombination factor is depicted in Fig.(\ref{S22x}).  
As shown in the figure, the stability of the black string solutions depends on the values of the magnetic charge ratios. For instance, in the case of $\tau = 100$, the recombination factor is consistently less than 1, indicating that only stable behaviors are observed. Conversely, for $\tau = 50$, both stable and unstable behaviors can occur depending on the specific value of  magnetic charges. It is worth noting that the variation in the recombination factor exhibits some apparent similarities in form across different cases.

\section{Conclusion and discussions}

In this paper, we  have   studied  certain physical behaviors of   black  branes 
from the compactification of M-theory on a four  parameter  CY threefold with  $h^{1,1} = 4$.  Combining toric geometry techniques  and  5D $ {\cal N} = 2$ supergravity
formalisms, we  have   approached  the BPS and  the non-BPS states of  the  balck holes and the  black strings  which are obtained  by wrapping M2 and M5-branes on
appropriate non-homomorphic  2-cycles and 4-cycles, respectively.  Concerning the  black holes,  we have found 
the allowed electric charge regions of the BPS and the non-BPS states using  the effective scalar potential computations.  Inspired by  extended black hole entropies, we have discussed  certain thermodynamic  behaviors  by computing the associated quantities including the entropy and   the  temperature. Then, we  have approached   the stability of the non-BPS black holes by computing  the recombination factor associated with four  CY K\"{a}hler   parameters.  For simplicity reasons,  we
have  treated only two solutions.  Analyzing  such branch
solutions, we have examined  the stability behaviors in the allowed electric charge regions of  5D M-theory  black holes. As expected, we have  found  stable and unstable non-BPS black hole  states
depending on the electric charge ratios.
After that,  we have investigated  the non-BPS black strings  that are obtained by wrapping  M5-branes
on  dual non-homormphic 4-cycles in the proposed  toric CY threefold. Precisely, we
have found multiple non-BPS solutions.  Using the extremization mechanism with respect to the geometric
local variables subject to the volume constraint, we have  elaborated  black string solutions from the expression of 
the stringy  effective scalar potential. In particular,  we have computed 
the recombination factor $R$.  After a close examination, we have shown that the associated non-BPS black string states are stable in  certain  allowed magnetic charge regions. In such regions, the 5D black strings do enjoy the recombination process.

In the end of this work, we could quote certain open questions which  may come up.  A classic one concerns    4D black  holes using  the M-theory compactifications. It would be  interesting to think about alternative compactifications  by implementing  manifolds with non-trivial features such as holonomy groups. $G_2$ manifolds could be a possible road to investigate   such classes of  black holes \cite{BSA,AB}.   Moreover, a  remarkable feature   of CY threefolds  is mirror symmetry playing a  fondamental role in the compactification scenarios of string theory and related topics including F-theory.  It would therefore be interesting to try to deploy such a geometric tool in CY black holes.  We believe that these  issues  deserve  more  reflections. We hope to address  such questions in future works.


\end{document}